\documentclass{ws-ijqi}

\usepackage{overcite}\let\refcite\citen\relax

\renewcommand{\draftnote}{\relax}
\renewcommand{\trimmarks}{\relax}

\newcommand{\PACS}[1]{\par\vspace*{8pt}%
{\authorfont{\leftskip18pt\rightskip\leftskip%
\noindent\textit{PACS}\/:\ #1\par}}\par}

\def\currenttime{\hour=\time \divide\hour by 60 \number\hour:%
  \multiply\hour by 60 \minute=\time \global\advance\minute by -\hour%
  \ifnum\minute<10 0\number\minute\else\number\minute\fi}

\newcommand{\JournalTitle}[1]{\textit{#1}\ }

\newcommand{\PRA}{\JournalTitle{Phys.\ Rev.\ A}}
\newcommand{\PRD}{\JournalTitle{Phys.\ Rev.\ D}}
\newcommand{\PRL}{\JournalTitle{Phys.\ Rev.\ Lett.}}
\newcommand{\PL}{\JournalTitle{Phys.\ Lett.}}
\newcommand{\ZPhys}{\JournalTitle{Z.~Phys.}}

\newcommand{\JPhysA}{\JournalTitle{J. Phys.\ A: Math.\ Gen.}}
\newcommand{\OC}{\JournalTitle{Opt.\ Commun.}}
\newcommand{\FP}{\JournalTitle{Found.\ Phys.}} 

\newcommand{\mc}[1]{\mathcal{#1}} 
\newcommand{\cP}{\mc{P}}
\newcommand{\cV}{\mc{V}}

\newcommand{\half}{\frac{1}{2}}
\newcommand{\thalf}{\tfrac{1}{2}}
\newcommand{\third}{\frac{1}{3}}
\newcommand{\tthird}{\tfrac{1}{3}}
\newcommand{\adj}{^{\dagger}}
\newcommand{\D}{\mathrm{d}}
\newcommand{\I}{\mathrm{i}}
\newcommand{\Exp}[1]{{\rm e}^{\mbox{\footnotesize$#1$}}}
\newcommand{\tr}[1]{\mathrm{tr}\left\{ #1 \right\}}

\newcommand{\trho}{\tilde{\varrho}}
\newcommand{\tp}{\tilde{p}}
\newcommand{\phstar}{^{\phantom{*}}}

\newcommand{\diag}[1]{\mathrm{diag}#1}
\newcommand{\offdiag}[1]{\mathrm{offdiag}#1}
\newcommand{\power}[1]{^{\mbox{\footnotesize$#1$}}}

\newcommand{\sech}{\mathop{\mathrm{sech}}}

\begin{document}

\markboth{B.-G.~Englert, D.~Kaszlikowski, L.~C.~Kwek, and W.~H.~Chee}
{Wave-particle duality in multi-path interferometers}

\title{\uppercase{Wave-particle duality in multi-path interferometers: General
  concepts and three-path interferometers}}

\author{\uppercase{Berthold-Georg Englert},$^{1,2,*}$ %
\uppercase{Dagomir Kaszlikowski},$^{1,2,\dag}$\\ %
\uppercase{Leong Chuan Kwek},$^{1-3,\ddag}$ %
and \uppercase{Wei Hui Chee}$^{1,\S}$}
\address{$^1$Department of Physics, %
National University of Singapore, Singapore 117542\\
$^2$Centre for Quantum Technologies, %
National University of Singapore, Singapore 117543\\
$^3$Nanyang Technological University, %
National Institute of Education, Singapore 259756\\ %
$^*$phyebg@nus.edu.sg\\ $^\dag$phykd@nus.edu.sg\\ 
$^\ddag$lckwek@nie.edu.sg,phyklc@nus.edu.sg\\ %
$^\S$yuniesan@yahoo.com.sg}

\maketitle

\begin{history}
\received{30 September 2007}
\end{history}

\begin{abstract}
For two-path interferometers, the which-path predictability $\cP$ 
and the fringe visibility $\cV$ are familiar quantities that are much 
used to talk about wave-particle duality in a quantitative way.
We discuss several candidates that suggest themselves as generalizations $P$
of $\cP$ for multi-path interferometers, and treat the case of three paths
in considerable detail.
To each choice for the \emph{path knowledge} $P$, 
the \emph{interference strength} $V$ 
--- the corresponding generalization of $\cV$ ---
is found by a natural, operational procedure.
In experimental terms it amounts to finding those equal-weight superpositions 
of the path amplitudes which maximize $P$ for the emerging intensities.   
Mathematically speaking, one needs to identify a certain optimal one among 
the Fourier transforms of the state of the interfering quantum object.
Wave-particle duality is manifest, inasmuch as $P=1$ implies $V=0$ and $V=1$
implies $P=0$, whatever definition is chosen.
The possible values of the pair $(P,V)$ are restricted to an area with
corners at $(P,V)=(0,0)$, $(P,V)=(1,0)$, and $(P,V)=(0,1)$, with the shape of
the border line from $(1,0)$ to $(0,1)$ depending on the particular choice
for $P$ and the induced definition of $V$.
\end{abstract}



\section{Introduction}
Einstein's wave-particle duality is arguably the most familiar phenomenon
resulting from Bohr's principle of complementarity.%
\footnote{This means a logical inference, not a historical one.
In fact, the historical order is reversed: first wave-particle duality
(1905) \cite{Einstein:05}, then complementarity (1928) \cite{Bohr:28}.}
The intense debate between these two protagonists, of which Bohr's
essay on the occasion of Einstein's 70th birthday is the best known public
record \cite{Bohr:49}, continues to be the object of scholarly
studies \cite{Held:98}, but there is, of course, rather wide-spread agreement 
by now on what used to be controversial issues then.

In the late 1920s and early 1930s, discussions of wave-particle duality
focused on the extreme situations where only the particle aspects are
present, or only the wave aspects. 
As natural as this focus may have been then, it does not do full justice to
the subject, as it ignores the intermediate situations in which both aspects
of an atomic system coexist within the boundaries that Nature imposes by the
laws of quantum mechanics.

The compromises that she permits are well understood in the context of
two-path interferometers, where various inequalities quantify to which extent
wave and particle properties can be observed simultaneously.%
\footnote{These matters are reviewed in Ref.~\refcite{Englert+1:00}, with a
summary of the history of the subject in which  
Refs.~\refcite{Wootters+1:79,Rauch+1:84,Glauber:86,Mittelstaedt+2:87,%
Greenberger+1:88,Mandel:91,Jaeger+2:95,Englert:96} play an
important role. 
A technically simpler account, perhaps to be recommended as a
first reading, is given in Ref.~\refcite{Englert:99}.} 
The historical example of the Bohr-Einstein debate --- Einstein's version of
Young's double-slit interferometer, with a recoiling first single slit ---
is familiar textbook material, but its literal realization has not been
achieved as yet. 

What has become experimental reality, however, 
are analogous two-path interferometers of various kinds: 
for photons \cite{Mittelstaedt+2:87,Baldzuhn+2:89,Baldzuhn+1:91,%
Schwindt+2:99,Kwiat+2:99,Trifonov+3:02,Walborn+3:02}, 
neutrons \cite{Rauch+1:84,Summhammer+2:87}, 
and atoms \cite{Durr+2:98a,Durr+2:98b,Mei+1:01}.
They enable quantitative studies of wave-particle duality, in
which the said inequalities are tested.%
\footnote{Some of the cited experiments are of a more qualitative nature,
however.} 
As expected, it is consistently found that the inequalities are obeyed, not
a single violation has been reported. 

The basic inequality reads
\begin{equation}
  \label{eq:P+V-bit}
  \cP^2+\cV^2\leq1\,;
\end{equation}
all others can be derived from it with more or less sophisticated arguments
\cite{Englert+1:00}.
Here, the \emph{predictability} $\cP$ quantifies the particle aspects:
the \textit{a priori\/} odds for guessing the path right are given by 
$\half(1+\cP)$, and  the \emph{visibility} $\cV$ is simply the 
standard fringe visibility, the quantitative measure for the wave aspect.

It is our primary objective in this paper to introduce, and discuss, the
generalization to multi-path interferometers, with a particular emphasis on
three-path interferometers where most features are already present in their
generic forms. 
To this end, we shall not employ D\"urr's strategy of Ref.~\refcite{Durr:01}, 
who aimed at generalizations of $\cP$ and $\cV$ such that the equal sign in
(\ref{eq:P+V-bit}) continues to apply for all pure states propagating through
an $n$-path interferometer, as it does for two-path interferometers.
Rather, we consider a few possible choices for a generalization of $\cP$ 
that suggest themselves and identify the corresponding generalization 
of $\cV$ in, so we think, a natural way. 

Our present effort is not the first of its kind. 
We have already mentioned D\"urr's work \cite{Durr:01}, which is quite
substantial, and an earlier discussion, rather brief and with no definite
conclusions, is contained in Ref.~\refcite{Jaeger+2:95}. 
We explore suggestions for generalizing $\cP$ from both. 
By contrast, the recent approach by Luis \cite{Luis:01}, who is motivated by
the experiment of Ref.~\refcite{Mei+1:01}, does not fit into our 
strategy; put tersely, he is concerned with ``this-path information'' whereas
we care for ``which-path information.'' 

Here is a brief outline.
After presenting our general strategy in Sec.~\ref{sec:General}, we
illustrate matters in the simple situation of two-path interferometers in
Sec.~\ref{sec:Bits}. 
This is followed by a detailed study of three-path interferometers in
Sec.~\ref{sec:Trits}, which exhibit the generic features of multi-path
interferometers. 
Four-path interferometers and multi-path interferometers are then briefly
dealt with in Sec.~\ref{sec:Quarts}. 
We close with a summary.

\begin{figure}[!t]
\centering\includegraphics{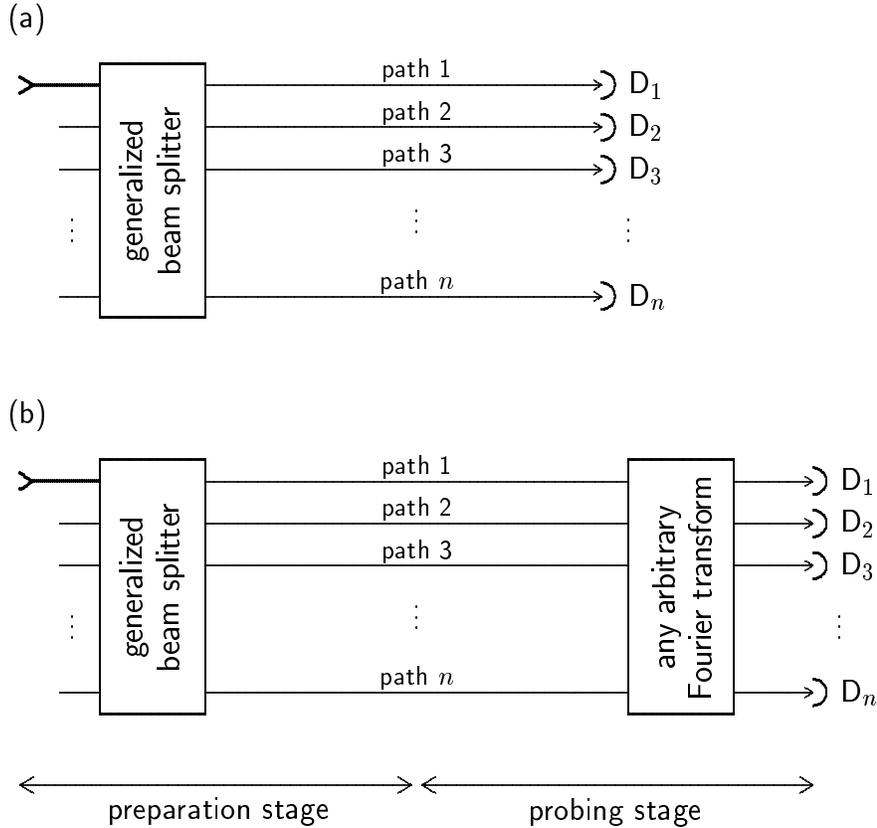}
\caption{\label{fig:nPaths}%
The two stages of a $n$-path interferometer: 
preparation stage and probing stage, 
and the two modes of operation: particle mode and wave mode. 
\textbf{Left side:} 
At the preparation stage, the incoming intensity (usually only one input
port is used) is distributed over all paths by a ``generalized beam splitter''
or \emph{\underline{n}itter}, which is the $n$-port version of the 
entry beam splitter of the common two-path interferometers.
A symmetric nitter  is unbiased and assigns equal intensity
to all paths (analogous to a symmetric 50:50 beam splitter), 
but it is more general to allow for a biased transformation (an asymmetric
beam splitter in the case of $n=2$), so that the intensity may vary from one
path to the next. 
\textbf{Right side:~(a)}  
In \emph{particle mode} (top), the probing just amounts to detecting the path,
a click of detector D$_m$ indicating that the $m$th path was the case.
\textbf{(b)}~In \emph{wave mode} (bottom),  
the probing stage uses a Fourier transformation, that is: a symmetric nitter, 
in front of the detectors, so that all paths contribute equally to the
intensity in each of the $n$ output ports.
In principle, any arbitrary Fourier transformation is to be considered, but in
practice a suitably chosen set of $n$ transformations suffice, each
characterized by the values of the relative phases between the amplitudes of
the paths.
The differences in the probabilities that the various detectors 
D$_1$, D$_2$, \dots, D$_n$ respond result only from these relative phases.
Taken together, the probabilities constitute the potentially complicated
interference pattern, in their dependence on those relative phases.
}
\end{figure}

\section{General considerations}\label{sec:General}
\subsection{Operating an interferometer in particle mode or wave mode}
\label{sec:InterModes}
In the most general terms, a $n$-path interferometer consists of an initial
preparation stage and a final probing stage, see Fig.~\ref{fig:nPaths}.
It is convenient to describe the state of the interfering system
between the stages by a $n\times n$ density matrix,
\begin{equation}
  \label{eq:A1}
  \varrho=\left(
    \begin{array}{cccc}
     \varrho_{11} & \varrho_{12} & \dots & \varrho_{1n} \\ 
     \varrho_{21} & \varrho_{22} & \dots & \varrho_{2n} \\
      \vdots & &\ddots& \vdots \\
     \varrho_{n1} & \multicolumn{2}{c}{\dots} & \varrho_{nn} 
    \end{array}\right)\,,\qquad\varrho\geq0\,,\qquad\tr{\varrho}=1\,.
\end{equation}
Of course, the diagonal entries $\varrho_{11}$, $\varrho_{22}$, \dots,
$\varrho_{nn}$  are the probabilities of finding the system in the 1st, 2nd,
\dots,  $n$th path, respectively. 
These probabilities are experimentally available by operating the
interferometer in the \emph{particle mode} of Fig.~\ref{fig:nPaths}(a). 

The \emph{wave mode} of Fig.~\ref{fig:nPaths}(b) probes the off-diagonal
elements in (\ref{eq:A1}).
The unitary $n\times n$ matrix $F$ of the Fourier transformation,
\begin{equation}
  \label{eq:A2}
  \varrho\to \trho=F\varrho F\adj\,,\qquad F\adj=F^{-1}\,,
\end{equation}
is such that all its matrix elements are of the same absolute size,
\begin{equation}
  \label{eq:A3}
  \left|F_{jk}\right|=\frac{1}{\sqrt{n}}\,,
\end{equation}
for which
\begin{equation}
  \label{eq:A4}
  F_{jk}=\left(F^{-1}\right)_{kj}^*
  =\frac{1}{\sqrt{n}}\Exp{\I2\pi jk/n}\Exp{\I\phi_j+\I\varphi_k}
\end{equation}
is the generic example, where $\phi_j$ and $\varphi_k$ are arbitrary phases.
If $\phi_j=0$ and $\varphi_k=0$ for all $j$ and $k$, we get the matrix of
the standard discrete Fourier transformation. 
The resulting probability of the $m$th detector to click is
\begin{equation}
  \label{eq:A5}
  \trho_{mm}=\left(F\varrho F\adj\right)_{mm}
   =\frac{1}{n}+\sum_{j\neq k}F_{mj}\phstar\varrho_{jk}F^*_{mk} 
\end{equation}
in general, and
\begin{equation}
  \label{eq:A6}
  \trho_{mm}=\frac{1}{n}+\frac{1}{n}\sum_{j\neq k}
\Exp{\I2\pi m(j-k)/n}\Exp{\I(\varphi_j-\varphi_k)}\varrho_{jk}
\end{equation}
in particular for the generic $F$ of (\ref{eq:A4}), where the phases $\phi_j$
are irrelevant in this context.

We must not fail to note that, as a consequence of the defining property
(\ref{eq:A3}) of the general Fourier transform, the two modes of operation in
Fig.~\ref{fig:nPaths} are complementary. 
For, if the path is certain in particle mode, that is: 
$\varrho_{mm}=\delta_{mm'}$ if the $m'$th path is the case, then all
detectors will click with equal probability in wave mode: $\trho_{mm}=1/n$. 
And conversely, if $\trho_{mm}=\delta_{mm'}$ for one Fourier transform
in Fig.~\ref{fig:nPaths}(b), then $\varrho_{mm}=1/n$ follows, so that all paths
are found with equal probability in Fig.~\ref{fig:nPaths}(a).

\subsection{Fourier matrices}\label{sec:Fourier}
What is hinted at in Eq.~(\ref{eq:A4}) can be carried out for any Fourier
matrix,
\begin{equation}
  \label{eq:F1}
  F=\left(\begin{array}{cccc}
     \Exp{\I\phi_1} & 0 & \dots & 0 \\ 
     0 & \Exp{\I\phi_2} & \dots & 0 \\
      \vdots & &\ddots& \vdots \\
     0 & \multicolumn{2}{c}{\dots} & \Exp{\I\phi_n} 
    \end{array}\right)
\frac{1}{\sqrt{n}}
\left(\begin{array}{c@{\quad}c@{\quad}cc}
     \cdot & \cdots & \cdots & 1 \\ 
     \vdots & \ddots &   & 1 \\
      \vdots & &\ddots& \vdots \\
     1 & 1 & \dots & 1 
    \end{array}\right)
\left(\begin{array}{cccc}
     \Exp{\I\varphi_1} & 0 & \dots & 0 \\ 
     0 & \Exp{\I\varphi_2} & \dots & 0 \\
      \vdots & &\ddots& \vdots \\
     0 & \multicolumn{2}{c}{\dots} & \Exp{\I\varphi_n}
    \end{array}\right)\,,
\end{equation}
where the input phases $\varphi_k$ and the output phases $\phi_j$ are pulled
out such that the central Fourier matrix has elements $1/\sqrt{n}$ in the
$n$th row and the $n$th column. 
Only $n-1$ of these $2n$ phases are relevant because the $\trho_{mm}$s do not
involve the output phases $\phi_k$, and the option to redefine
all phases jointly in accordance with
\begin{equation}
  \label{eq:F2}
  \phi_j\to\phi_j+\alpha\,,\quad \varphi_k\to\varphi_k-\alpha
\qquad\mbox{($\alpha$ arbitrary)}
\end{equation}
can be used to set, say, $\varphi_n=0$ by convention. 

For $n=2$ this gives a unique central Fourier matrix,
\begin{equation}
  \label{eq:F3}
  F=\frac{1}{\sqrt{2}}
\left(\begin{array}{rc}-1&1\\1&1\end{array}\right)\,,
\end{equation}
and for $n=3$ there are two possible central Fourier matrices,
\begin{equation}
  \label{eq:F4}
  F=\frac{1}{\sqrt{3}}
\left(\begin{array}{ccc}\Exp{\I2\pi/3}&\Exp{-\I2\pi/3}&1\\
\Exp{-\I2\pi/3}&\Exp{\I2\pi/3}&1\\1&1&1\end{array}\right)
\quad\mbox{and}\quad
  F=\frac{1}{\sqrt{3}}
\left(\begin{array}{ccc}\Exp{-\I2\pi/3}&\Exp{\I2\pi/3}&1\\
\Exp{\I2\pi/3}&\Exp{-\I2\pi/3}&1\\1&1&1\end{array}\right)\,,
\end{equation}
but these two are equivalent for our purposes because they differ only by a
permutation of rows, that is: of the output channels, which can be compensated
for by a relabeling of the detectors in Fig.~\ref{fig:nPaths}(b).

For $n=4$, we have a one-parametric family of possible central Fourier
matrices,
\begin{equation}
  \label{eq:F5}
   F=\frac{1}{2}
\left(\begin{array}{crcr}
 \phantom{-}\Exp{\I t}&-1&-\Exp{\I t}&1 \\ -1&1&-1&1\\
 -\Exp{\I t}&-1& \phantom{-}\Exp{\I t}&1 \\
 \phantom{-}1&1& \phantom{-}1&\enskip1\end{array}\right)
\qquad\mbox{with arbitrary real $t$}\,,
\end{equation}
supplemented by those matrices that one obtains by permutations of columns
that cannot be undone by permuting rows.
The Fourier matrix of (\ref{eq:A4}) corresponds to $t=\pi/2$, and $t=0$ in
conjunction with permuting the 2nd and 3rd rows gives the tensor product of 
the $2\times2$ matrix in (\ref{eq:F3}) with itself.

For $n=5$, the situation is similar to that for $n=3$, as there is essentially
only one central Fourier matrix, the standard one of (\ref{eq:A4}).
Unfortunately, this is not true for other prime values of $n$. 
For example, there are five inequivalent choices for $n=7$.
And for composite values of $n$, we have continuous families of central
Fourier matrices, is illustrated above for $n=4=2\times2$;
more about this at the website of Ref.~\refcite{Zyczkowski+1:website}.

Our choice of terminology to refer to all matrices with the property
(\ref{eq:A3}) as \emph{Fourier matrices} is not everybody's convention.
Some authors speak of \emph{Hadamard matrices} instead,%
\footnote{By convention, Fourier matrices are unitary, $F^{-1}=F^\dagger$, 
whereas Hadamard matrices are normalized to unit-modulus matrix elements, 
such that $nH^{-1}=H^\dagger$, and corresponding matrices are
related by $H=\sqrt{n}F$.}
thus generalizing the real Hadamard matrices of combinatorics 
which have $\pm1$ as matrix elements  
--- such as the $2\times2$ matrix in (\ref{eq:F3}) or the $t=0$ version of the
$4\times4$ matrix in (\ref{eq:F5}) --- to complex matrices, and our
\emph{central} Fourier matrices of (\ref{eq:F1}) are called \emph{dephased}
Hadamard matrices. 
Unfortunately, the general parameterization of all Fourier or Hadamard
$n\times n$ matrices is not known for arbitrary values of $n$.
A concise guide is Ref.~\refcite{Tadej+1:06}, 
and a catalog of known cases up to $n=16$ is available at
the web site maintained, in a truly commendable effort, by \.Zyczkowski 
and Tadej,\cite{Zyczkowski+1:website} where one also finds an extensive 
list of references on the subject.

From the point of view of quantum physics, the elements of a Fourier matrix
are the transition amplitudes between two mutually unbiased bases. 
Accordingly, the particle-mode and the wave-mode operation of the $n$-path
interferometer in Fig.~\ref{fig:nPaths} realize the measurements of a pair of
complementary observables, as we noted at the end of Sec.~\ref{sec:InterModes}.
In this context one usually encounters Fourier transformations, and this
prompted our choice of terminology.

\subsection{Quantification of the path knowledge}\label{sec:particleness}
Path knowledge is knowledge about the probabilities for detector
clicks in Fig.~\ref{fig:nPaths}(a), that is: knowledge about the diagonal
elements of the density matrix $\varrho$ in (\ref{eq:A1}).
In view of the normalization of $\varrho$ to unit trace, one needs $n-1$ real
parameters to specify all $\varrho_{mm}$.
For example, the real and imaginary
parts of the complex numbers $z_1$, $z_2$, \dots, $z_{n-1}$ that are defined by
\begin{equation}
  \label{eq:A7}
  z_k\phstar=\sum_{m=1}^n\Exp{\I2\pi km/n}\varrho_{mm}=z_{n-k}^*
\end{equation}
may serve this purpose.
Clearly, then, there cannot be a unique universal way of quantifying path 
knowledge by a single number (except for $n=2$), and various numerical 
measures will be justifiable. 
To a considerable extent, it thus remains a matter of taste, or convenience,
for which of them to opt, unless particular circumstances leave no choice.
     
We shall regard any continuous function 
$P(\diag{\varrho})\equiv P(\varrho_{11},\varrho_{22},\ldots,\varrho_{nn})$
of the diagonal elements of $\varrho$ as an acceptable measure of path
knowledge, and thus as a valid generalization of the two-path 
predictability $\cP$, if it meets these natural criteria:
\begin{equation}
  \label{eq:A8}
\parbox{0.78\columnwidth}{%
\makebox[0pt][r]{a.~}$P=1$ if $\varrho_{mm}=1$ for one
$m$, i.e., if the path is certain, and only then.\\
\makebox[0pt][r]{b.~}$P=0$ if $\varrho_{mm}=1/n$ for all
$m$, i.e., if the path is completely uncertain.\\
\makebox[0pt][r]{c.~}$P$ must be invariant under
permutations of the diagonal elements of $\varrho$.\\
\makebox[0pt][r]{d.~}$P$ must be convex, that is:
\begin{displaymath}
P(\diag{\varrho})\leq(1-\lambda) P(\diag{\varrho_1})
+\lambda P(\diag{\varrho_2})
\end{displaymath}
with $\varrho=(1-\lambda)\varrho_1+\lambda\varrho_2$ 
and $0\leq\lambda\leq1$ holds  
for any two density
matrices $\varrho_1$ and $\varrho_2$.\\
\makebox[0pt][r]{e.~}Any degradation of the $\varrho_{mm}$s, that is: the
increase of a smaller one at the expense of a larger one, must not increase
the value of $P$. 
}
\end{equation}
Property (\ref{eq:A8}e) is actually implied by property (\ref{eq:A8}d), 
but we list it nevertheless because it is a weaker version of D\"urr's 
fourth criterion, at Eq.~(1.14) in Ref.~\refcite{Durr:01}, which requires 
``should decrease'' rather than ``must not increase.''   
The other properties are equivalent to D\"urr's.

\subsubsection{First example: Betting on the path}\label{sec:P-bet}
At Eq.~(\ref{eq:P+V-bit}), we recalled that the standard predictability $\cP$ 
of two-path interferometers is essentially the odds for guessing the way
right.  
More generally, then, a path-knowledge function $P(\diag{\varrho})$ can be
associated with a given  set of betting rules, and this will serve as our
first example. 

Whereas there is really only one kind of bet for $n=2$, there is a variety of
possible bets in $n$-path interferometers.
We consider bets of the following construction.
 \begin{equation}
  \label{eq:A9}
\parbox{0.80\columnwidth}{%
If you guess the path right on the 1st try, you win $g_1=1$ unit.
If your 1st guess is wrong, but your 2nd is right,
you win $g_2$ units.
And so forth: If you need $m$ guesses, you win $g_m$ units, and if all your   
$n-1$ guesses are wrong, you win $g_n$ units (which will be a negative
amount, so that you actually lose).
}
\end{equation}
The amount won should be the larger, the fewer guesses you need, and a random
guess should have a neutral over-all return.
These natural requirements impose the restrictions
\begin{equation}
  \label{eq:A10}
 1=g_1> g_2 \geq g_3 \geq \ldots \geq g_n\,,\qquad\sum_{m=1}^ng_m=0\,.
\end{equation}

The optimal betting strategy is clearly to first bet on the most likely path,
then on the second likely, and so on.
On average, the gain is then
\begin{equation}
  \label{eq:A11}
  P_\mathrm{bet}=\sum_{m=1}^ng_mp_m\,,
\end{equation}
where the $p_m$s are the $\varrho_{mm}$s in descending order,
\begin{equation}
  \label{eq:A12}
  \bigl\{p_1,p_2,\dots,p_n\bigr\}
=\bigl\{\varrho_{11},\varrho_{22},\dots,\varrho_{nn}\bigr\}\,,\qquad
  p_1\geq p_2 \geq \ldots \geq p_n\geq0\,.
\end{equation}
By construction, $P_\mathrm{bet}$ meets the criteria (\ref{eq:A8}a--e), 
and so $P_\mathrm{bet}$ is an acceptable numerical measure of path knowledge.
It is, in fact, the relevant measure if the bet specified by the choice 
of $g_m$s is the operational procedure for verifying someone's claim 
that he has such knowledge.

A particularly simple case is the ``one-guess bet,'' specified by
$g_1=1$, $g_2=g_3=\cdots=g_n=-1/(n-1)$, for which
\begin{eqnarray}
  \label{eq:A13}
 P_\mathrm{bet}^{\mathrm{(1\,guess)}}&=&\frac{n}{n-1}p_1-\frac{1}{n-1}
\nonumber\\
&=&\frac{n}{n-1}\max_m\bigl\{\varrho_{mm}\bigr\}-\frac{1}{n-1}.
\end{eqnarray}
This is the proposal that is briefly discussed in Appendix~C of
Ref.~\refcite{Jaeger+2:95}.
Equally natural is the ``linear bet,''
\begin{equation}
  \label{eq:A14}
   P_\mathrm{bet}^{\mathrm{(lin)}}
=\frac{n+1}{n-1}-\frac{2}{n-1}\sum_{m=1}^nmp_m\;,
\end{equation}
that has $g_m=(n+1-2m)/(n-1)$.
Harkening back to the remark after (\ref{eq:A8}), we note that the linear bet
meets D\"urr's stronger requirement that any degradation should decrease
$P$, whereas the one-guess bet does so only for $n=2$.

\subsubsection{Second example: Normalized purity}\label{sec:P-Durr}
As a second example, we consider D\"urr's proposal of Ref.~\refcite{Durr:01},
who constructs a path-knowledge function $P(\diag{\varrho})$ from the
so-called ``purity'' of the probability distribution, essentially the sum of 
the squared path probabilities.
D\"urr's path-knowledge function is
\begin{equation}
  \label{eq:A15}
  P_\mathrm{pur}=\left(\frac{n}{n-1}\sum_{m=1}^n\varrho_{mm}^2
                      -\frac{1}{n-1}\right)\power{\half}\,,
\end{equation}
which is properly normalized to meet requirements (\ref{eq:A8}a) and
(\ref{eq:A8}b).

\subsubsection{Third example: Normalized Shannon entropy}\label{sec:P-ent}
From the Shannon entropy\cite{Shannon:48} associated with 
the probability distribution $\diag{\varrho}$,
\begin{equation}
  \label{eq:A16}
  S(\diag{\varrho})=-\sum_{m=1}^n\varrho_{mm}\log\varrho_{mm}
\end{equation}
one can construct yet another path-knowledge function.
This approach was regarded as the natural one by the authors of
Refs.~\refcite{Wootters+1:79} and \refcite{Mittelstaedt+2:87} in the context
of two-path interferometers, but was found less appealing for $n$-path
interferometers by the authors of Refs.~\refcite{Jaeger+2:95} and
\refcite{Durr:01}.  

Upon proper normalization to meet requirements (\ref{eq:A8}a) and
(\ref{eq:A8}b), the entropic measure of path knowledge is given by
\begin{equation}
  \label{eq:A17}
  P_\mathrm{ent}
=\frac{1}{\log n}\sum_{m=1}^n\varrho_{mm}\log(n\varrho_{mm})\,.
\end{equation}
Whereas the binary logarithm is usually understood in (\ref{eq:A16}), 
it does not matter which base value is chosen in  (\ref{eq:A17}).

\subsubsection{Fourth example: R\'enyi-type measures}\label{sec:P-Ren}
It is worth mentioning that, as a generalization of both the purity measure in
(\ref{eq:A15}) and the entropic measure in (\ref{eq:A17}), one could employ the
R\'enyi-type measures that are defined by
\begin{equation}
  \label{eq:A17a}
  P_\mathrm{Ren}^{(\lambda)}=\left(\frac{n^\lambda}{n^\lambda-n}
          \sum_{m=1}^n\varrho_{mm}^\lambda
                      -\frac{n}{n^\lambda-n}\right)\power{\frac{1}{\lambda}}
\end{equation}
where $\lambda$ is a positive parameter.%
\footnote{The symbol $p$ is usually used to denote the parameter in the
standard definition of the family of R\'enyi entropies, but we switch to
$\lambda$ in order to avoid confusion with the probabilities $p_1, \dots, p_n$.
Further we note that we find it convenient to not take the logarithm of the
sum in (\ref{eq:A17a}), which would be yet another option.}
We recover the purity measure for $\lambda=2$ and the entropic measure in the
limit $\lambda\to1$,
\begin{equation}
  \label{eq:A17b}
  P_\mathrm{Ren}^{(2)}=P_\mathrm{pur}\,,\qquad
  P_\mathrm{Ren}^{(1)}\equiv P_\mathrm{Ren}^{(\lambda\to1)}=P_\mathrm{ent}\,,
\end{equation}
and intermediate $\lambda$ values interpolate between $P_\mathrm{pur}$ and
$P_\mathrm{ent}$.

There are also the limits $\lambda\to\infty$ and $\lambda\to0$, both of which
are peculiar. 
We have
\begin{equation}
  \label{eq:A17c}
   P_\mathrm{Ren}^{(\infty)}\equiv P_\mathrm{Ren}^{(\lambda\to\infty)}=
\left\{
  \begin{array}{c@{\mbox{\enskip if\enskip}}l}
   p_1 & \displaystyle np_1>1\,,\\[1ex]
    0  & \displaystyle np_1=1\,,      
    \end{array}\right.
\end{equation}
for $\lambda\to\infty$, and 
\begin{equation}
  \label{eq:A17d}
   P_\mathrm{Ren}^{(0)}\equiv P_\mathrm{Ren}^{(\lambda\to0)}=
\left\{
  \begin{array}{c@{\mbox{\enskip if\enskip}}l}
    1 & p_1=1\,,\\[1ex]
    0  & p_1<1\,,      
    \end{array}\right.
\end{equation}
for $\lambda\to0$.
For both, the value of $\displaystyle p_1=\max_m\{\varrho_{mm}\}$ matters
solely, which makes $P_\mathrm{Ren}^{(\infty)}$, and to a much lesser extent also
$P_\mathrm{Ren}^{(0)}$, somewhat similar to $P_\mathrm{bet}^{\mathrm{(1\,guess)}}$.

We are not examining these R\'enyi-type measures in the situations of $n=2$
and $n=3$ that are dealt with in Secs.~\ref{sec:Bits} and \ref{sec:Trits}, but
will offer a few comments on the limiting measures $P_\mathrm{Ren}^{(\infty)}$ and
$P_\mathrm{Ren}^{(0)}$ for arbitrary $n$ values in Sec.~\ref{sec:Nits}.

\subsection{Quantification of the tendency for interference}
\label{sec:waveness}
Rather than introducing independent numerical measures for the strength of
the interference between the paths, we derive the corresponding wave quantity
from the given particle quantity, that is: the given path-knowledge function
$P(\diag{\varrho})$. 
Since we are thus constructing generalizations of the two-path visibility
$\cV$, the letter $V$ will be used for these measures of the tendency for
interference. 

After the Fourier transformation in Fig.~\ref{fig:nPaths}(b), we have the
density matrix $\trho=F\varrho F\adj$ and $P(\diag{\trho})$ tells us
how much path knowledge is available in the transformed state.
As the explicit expression (\ref{eq:A5}) for the diagonal elements of $\trho$
shows, the value of $P(\diag{\trho})$ depends crucially on the off-diagonal
elements of density matrix $\varrho$, but not at all on the diagonal elements
that yield the value of $P(\diag{\varrho})$. 
Further, the particular choice for $F$ enters $P(\diag{\trho})$, 
and so its value 
will be rather small for some Fourier transformations and particularly large 
for others. 

Accordingly, with the intention of quantifying the joint size of the
off-diagonal elements of $\varrho$ in a fitting manner, we define 
$V(\offdiag{\varrho})$ as the largest value that $P(\diag{\trho})$ can attain,
\begin{equation}
  \label{eq:A18}
  V(\offdiag{\varrho})=\max_{F}P(\diag{\left\{F\varrho F\adj\right\}})\,,
\end{equation}
where the maximum is sought in the set of all Fourier matrices $F$, that is:
all unitary matrices that obey (\ref{eq:A3}).
Harkening back to the remark at the end of Sec.~\ref{sec:Fourier}, we note
that the maximization in (\ref{eq:A18}) is over all measurements of
observables that are complementary to the path observable of
Fig.~\ref{fig:nPaths}(a). 
Accordingly, this way of quantifying the interference strength has an
unambiguous operational meaning.

As a consequence of the optimization in (\ref{eq:A18}),
the properties (\ref{eq:A8}) of $P(\diag{\varrho})$ have their counterparts
for $V(\offdiag{\varrho})$, namely 
\begin{equation}
  \label{eq:A19}
\parbox{0.78\columnwidth}{%
\makebox[0pt][r]{a.~}$V=0$ if $\varrho_{mm}=1$ for one
$m$, i.e., if the path is certain.\\
\makebox[0pt][r]{b.~}$V=1$ is only possible if $\varrho_{mm}=1/n$ for all
$m$, i.e., if the path is completely uncertain, and only then.\\
\makebox[0pt][r]{c.~}$V$ is invariant under
permutations of the path labels.\\
\makebox[0pt][r]{d.~}$V$ is convex, that is:
\begin{displaymath}
V(\offdiag{\varrho})\leq(1-\lambda) V(\offdiag{\varrho_1})%
+\lambda V(\offdiag{\varrho_2})
\end{displaymath}
with $\varrho=(1-\lambda)\varrho_1+\lambda\varrho_2$ 
and $0\leq\lambda\leq1$ holds
for any two density matrices $\varrho_1$ and $\varrho_2$.\\
\makebox[0pt][r]{e.~}A degradation of the $\varrho_{jk}$s ($j\neq k$), 
that is: a reduction in size, cannot increase the value of $V$. 
}
\end{equation}
The maximum in (\ref{eq:A18}) is crucial in establishing the convexity
(\ref{eq:A19}d); if instead we took a single Fourier matrix $F$ in
(\ref{eq:A18}), or a too-small subset of Fourier matrices, the resulting
interference-strength measure $V$ would not be convex and, therefore, of
rather limited use.  

Clearly, then, for each set of betting rules in Sec.~\ref{sec:P-bet}
there is an interference-strength measure $V_\mathrm{bet}$ derived from the
corresponding path-knowledge measure $P_\mathrm{bet}$.
In particular, we have $V_\mathrm{bet}^{\mathrm{(1\,guess)}}$
and $V_\mathrm{bet}^{\mathrm{(lin)}}$ paired with 
$P_\mathrm{bet}^{\mathrm{(1\,guess)}}$ of (\ref{eq:A14})
and $P_\mathrm{bet}^{\mathrm{(lin)}}$ of (\ref{eq:A15}), respectively.
And likewise, we have $V_\mathrm{pur}$ associated with D\"urr's 
$P_\mathrm{pur}$ of Sec.~\ref{sec:P-Durr}, 
and also an entropic $V_\mathrm{ent}$ that goes
with $P_\mathrm{ent}$ of Sec.~\ref{sec:P-ent}. 

Each $P,V$ pair can be used to study the compromises intermediate between the
extreme situations of ``particle aspect only'' ($P=1$ and $V=0$) and ``wave
aspect only'' ($V=1$ and $P=0$). 
Qualitatively, the same picture emerges for all $P,V$ pairs: 
Owing to the complementarity of the particle and wave modes of operation
(recall the remark at the end of Sec.~\ref{sec:InterModes}), the two aspects
are mutually exclusive --- if $P=1$, then surely $V=0$, and vice versa.
In a $P,V$ diagram, the extremal points $(P,V)=(1,0)$ and $(P,V)=(0,1)$ are
connected by a (curved) line, which together with the straight lines to
$(P,V)=(0,0)$ encloses the area of all possible pairs of $P,V$ values.
The shape of the line from $(P,V)=(1,0)$ to $(P,V)=(0,1)$ and other
quantitative details depend on the specific choice for the path-knowledge
function $P(\diag\varrho)$ and the induced interference-strength measure
$V(\offdiag\varrho)$.

\section{Two-path interferometers: Qubits}\label{sec:Bits} 
In Sec.~\ref{sec:General}, we have emphasized consistently, and somewhat
pedantically, that the various path-knowledge functions $P(\diag{\varrho})$ 
involve only the diagonal elements of $\varrho$, and the corresponding 
interference-strength measures $V(\offdiag{\varrho})$ only the off-diagonal 
elements.
But from now on we shall simplify the notation and just write 
$P(\varrho)$ and $V(\varrho)$.

For a first illustration, and to make contact with familiar notions, we now
consider the particularly simple situation of two-path interferometers.
Then, the interfering object constitutes a binary quantum alternative, or
\emph{qubit}.%
\footnote{\label{fn:4}The alternative spelling \emph{q-bit} is less popular. 
One of us, at least, thinks that this is regrettable, 
because ``q-bit'' perfectly fits the pattern of Dirac's 
``q-numbers'' and ``c-numbers.''\label{fn:spelling}}
The familiar expressions for the predictability $\cP$ and the visibility
$\cV$,
\begin{equation}
  \label{eq:B1}
  \cP=\bigl|\varrho_{11}-\varrho_{22}\bigr|\,,\qquad
  \cV=2\bigl|\varrho_{12}\bigr|\,,
\end{equation}
are quite simply related to the elements of the $2\times2$ density matrix,
\begin{equation}
  \label{eq:B2}
  \varrho_\mathrm{bit}=\left(
    \begin{array}{cc}
    \varrho_{11} & \varrho_{12} \\ \varrho_{21} & \varrho_{22} 
    \end{array}\right)\,,
\end{equation}
and the basic inequality (\ref{eq:P+V-bit}) is an immediate consequence of the
normalization of $\varrho_\mathrm{bit}$ to unit trace
($\varrho_{11}+\varrho_{22}=1$) and its positivity
($\varrho_{12}\varrho_{21}\leq\varrho_{11}\varrho_{22}$).

The $n=2$ versions of the path-knowledge measures introduced in 
Sec.~\ref{sec:particleness} are simple monotonic functions 
of the predictability,
\begin{eqnarray}
  \label{eq:B3}
  && P_\mathrm{bet}=P_\mathrm{pur}=\cP\,,\nonumber\\[1ex]
  && P_\mathrm{ent}=\frac{1}{\log4}
           \bigl[(1+\cP)\log(1+\cP)+(1-\cP)\log(1-\cP)\bigr]\,,
\end{eqnarray}
and the implied interference-strength measures of Sec.~\ref{sec:waveness} are
the same functions of the visibility,
\begin{eqnarray}
  \label{eq:B4}
  && V_\mathrm{bet}=V_\mathrm{pur}=\cV\,,\nonumber\\[1ex]
  && V_\mathrm{ent}=\frac{1}{\log4}
           \bigl[(1+\cV)\log(1+\cV)+(1-\cV)\log(1-\cV)\bigr]\,.
\end{eqnarray}
For each of these $P,V$ pairs, pure states maximize the value of $P$ for
given $V$, and the value of $V$ for given $P$.
This observation is an immediate consequence of (\ref{eq:P+V-bit}),
where the equal sign applies for pure states. 

\begin{figure}[!t]
\centering\includegraphics{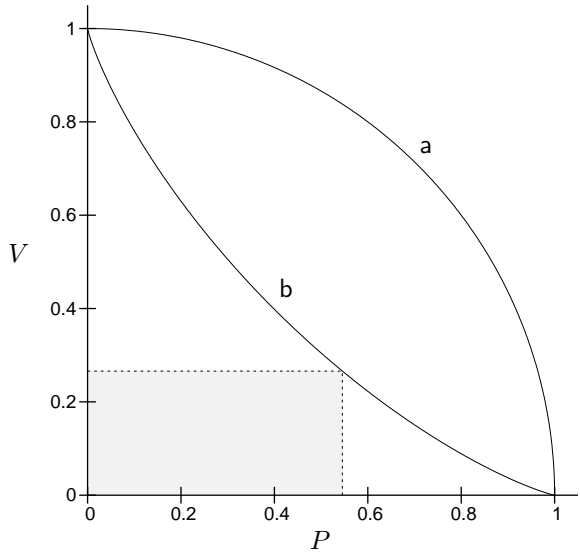}
\caption{\label{fig:qubit}%
Possible values for the path-knowledge measure $P$ and the
interference-strength measure $V$ in two-path interferometers.
Curve \textsf{a} is the quarter-circle border line for 
$(P,V)=(P_\mathrm{bet},V_\mathrm{bet})$ and
$(P,V)=(P_\mathrm{pur},V_\mathrm{pur})$, 
curve \textsf{b} is the border line for
$(P,V)=(P_\mathrm{ent},V_\mathrm{ent})$. 
Pure-state values are on the respective border curves, mixed-state values are
inside the area with corners at $(P,V)=(0,0)$, $(P,V)=(1,0)$, 
and $(P,V)=(0,1)$.
The shaded rectangle has the top-right corner on curve \textsf{b}; see text.}
\end{figure}

To trace the border curve that connects $(P,V)=(1,0)$ with $(P,V)=(0,1)$ it
is quite sufficient to consider the projector matrices
\begin{equation}
  \label{eq:B5}
    \varrho_\mathrm{bit,pure}=\left(
    \begin{array}{c}
    \cos\vartheta \\ \sin\vartheta
    \end{array}\right)\bigl( \cos\vartheta,\sin\vartheta\bigr)
\qquad\mbox{with\ } 0\leq\vartheta\leq\frac{1}{4}\pi
\end{equation}
because any other pure-state density matrix differs from one of these
$ \varrho_\mathrm{bit,pure}$s at most by a permutation and a 
phase transformation, which are irrelevant in the present context. 
For both the `bet' pair and the `pur' pair, the border curve is the quarter
circle labeled \textsf{a} in Fig.~\ref{fig:qubit}.
For the `ent' pair, the border is drawn by the concave curve \textsf{b}, 
which appears to be at odds with the convexity of $P_\mathrm{ent}$ 
and $V_\mathrm{ent}$ but in fact is not.

To justify this assertion we consider the qubit in an arbitrary mixed state,
so that 
\begin{equation}
  \label{eq:B6}
\varrho_{11}\varrho_{22}>0\qquad\mbox{and}\qquad
  \varrho_{12}=\varepsilon\sqrt{\varrho_{11}\varrho_{22}}
\quad\mbox{with\ }|\varepsilon|<1
\end{equation}
in (\ref{eq:B2}).
Now write $\varepsilon=|\varepsilon|\Exp{\I\alpha}$ and 
define $\lambda=\half(1+|\varepsilon|)$.
Then
\begin{eqnarray}
  \label{eq:B7}
  \varrho_\mathrm{bit}&=&\lambda\left(
    \begin{array}{cc}
    \varrho_{11} & \Exp{\I\alpha}\sqrt{\varrho_{11}\varrho_{22}} 
    \\  \Exp{-\I\alpha}\sqrt{\varrho_{11}\varrho_{22}}  & \varrho_{22} 
    \end{array}\right)
+(1-\lambda)\left(
    \begin{array}{cc}
    \varrho_{11} & -\Exp{\I\alpha}\sqrt{\varrho_{11}\varrho_{22}} 
    \\  -\Exp{-\I\alpha}\sqrt{\varrho_{11}\varrho_{22}}  & \varrho_{22} 
    \end{array}\right)
\nonumber\\[1ex]&\equiv& 
  \lambda\varrho_\mathrm{bit}^{(1)}+(1-\lambda)\varrho_\mathrm{bit}^{(2)}\,.
\end{eqnarray}
The two pure-state density matrices thus introduced, 
$\varrho_\mathrm{bit}^{(1)}$ and $\varrho_\mathrm{bit}^{(2)}$, have the same
predictability and visibility, namely 
$\cP=\bigl|\varrho_{11}-\varrho_{22}\bigr|$ and
$\cV=2\sqrt{\varrho_{11}\varrho_{22}}$.
Therefore, they also have the same $P,V$ pair of values, 
irrespective of whether we chose the `bet' pair, or the `pur' pair, or the
`ent' pair.
The convexity of $P(\varrho)$ and $V(\varrho)$ then implies
\begin{equation}
  \label{eq:B8}
  P(\varrho_\mathrm{bit})\leq P^{(1,2)}\,,\qquad
  V(\varrho_\mathrm{bit})\leq V^{(1,2)}\,,
\end{equation}
where $P^{(1,2)}$ and $V^{(1,2)}$ are the common values of
$\varrho_\mathrm{bit}^{(1)}$  and $\varrho_\mathrm{bit}^{(2)}$.
This says that, in Fig.~\ref{fig:qubit}, the point 
$\bigl(P(\varrho_\mathrm{bit}),V(\varrho_\mathrm{bit})\bigr)$ is inside the
rectangle with ${0\leq P\leq P^{(1,2)}}$, ${0\leq V\leq V^{(1,2)}}$.%
\footnote{Actually we have $P(\varrho_\mathrm{bit})=P^{(1,2)}$ so that point
$\mbox{\normalsize(}P(\varrho_\mathrm{bit}),%
V(\varrho_\mathrm{bit})\mbox{\normalsize)}$ lies on the
right border of the rectangle, but this detail is not relevant for the
argument.}\  
For an exemplary pair  $P^{(1,2)},V^{(1,2)}$ on curve \textsf{b},
this rectangle is  shaded in Fig.~\ref{fig:qubit}.
Clearly, it is wholly inside the area bounded by the axes and 
curve~\textsf{b}.

\section{Three-path interferometers: Qutrits}\label{sec:Trits}
Somewhat more interesting than two-path interferometers are three-path
interferometers, in which the interfering object is a ternary quantum
alternative, or \emph{qutrit}.%
\footnote{Footnote \ref{fn:4} applies \textit{mutatis mutandis}.}  
Here we have a $3\times3$ density matrix
\begin{equation}
  \label{eq:C1}
   \varrho_\mathrm{trit}^{\ }=\left(
    \begin{array}{ccc}
    \varrho_{11} & \varrho_{12} & \varrho_{13} \\ 
    \varrho_{21} & \varrho_{22} & \varrho_{23} \\
    \varrho_{31} & \varrho_{32} & \varrho_{33} 
    \end{array}\right)
\end{equation}
that is normalized to unit trace ($\varrho_{11}+\varrho_{22}+\varrho_{33}=1$)
and positive, so that the restrictions
\begin{equation}
  \label{eq:C2}
\tr{\varrho_\mathrm{trit}^2}\leq1\,,\quad
\det\bigl\{\varrho_\mathrm{trit}^{\ }\bigr\}\geq0
\end{equation}
apply.
Since phase transformations,
\begin{equation}
  \label{eq:C3}
  \varrho_{jk}\phstar\to\Exp{\I(\varphi_j-\varphi_k)}\varrho_{jk}\phstar\,,
\end{equation}
turn a given $\varrho_\mathrm{trit}^{\ }$ into an equivalent one, we can
adjust the complex phases of the off-diagonal elements such that 
\begin{equation}
  \label{eq:C4}
  \varrho_{12}\phstar=\left|\varrho_{12}\phstar\right|
\Exp{\third\I\theta}\,,\quad
  \varrho_{23}\phstar=\left|\varrho_{23}\phstar\right|
\Exp{\third\I\theta}\,,\quad
  \varrho_{31}\phstar=\left|\varrho_{31}\phstar\right|\Exp{\third\I\theta}\,,
\end{equation}
with a common phase factor $\exp(\third\I\theta)$ that, for the given
off-diagonal elements, is determined by the phase-invariant product
\begin{equation}
  \label{eq:C5}
 \varrho_{12}\phstar\varrho_{23}\phstar\varrho_{31}\phstar=
\bigl|\varrho_{12}\phstar\varrho_{23}\phstar\varrho_{31}\phstar\bigr|
 \Exp{\I\theta}\,.
\end{equation}
With the convention that $\theta=0$ if
$\varrho_{12}\phstar\varrho_{23}\phstar\varrho_{31}\phstar=0$  
and $-\pi<\theta\leq\pi$ otherwise, the value of $\theta$ is unique. 

For $n=3$, there is essentially only one complex number in the set of 
Eq.~(\ref{eq:A7}), namely $z_1\phstar=z_2^*\equiv z$, explicitly given by
\begin{equation}
  \label{eq:C6}
  z=q\varrho_{11}+q^2\varrho_{22}+\varrho_{33}\,,
\end{equation}
where
\begin{equation}
  \label{eq:C7}
  q\equiv\Exp{\I2\pi/3}=\frac{-1+\I\sqrt{3}}{2}
\end{equation}
is the basic cubic root of unity, for which
\begin{equation}
  \label{eq:C8}
  q^3=1\,,\quad q^2=q^*=q^{-1}\,,\quad 1+q+q^2=0 
\end{equation}
are noteworthy identities.
One can regard $z$ as the average of $q$, $q^2$, and $q^3=1$ that refers to
the weights $\varrho_{11}$, $\varrho_{22}$, $\varrho_{33}$, respectively.
Accordingly, when represented as points in Gauss's complex plane,
the possible values of $z$ are inside the equilateral triangle 
that has its corners at $1$, $q$, and $q^2$.

The identities (\ref{eq:C8}) are used, for example, 
when expressing the diagonal elements of 
$\varrho_\mathrm{trit}$ in terms of $z$,
\begin{equation}
  \label{eq:C9}
\left.
  \begin{array}{rl}
  \varrho_{11}&=\tthird(1+q^2z+qz^*)\\[0.8ex]
  \varrho_{22}&=\tthird(1+qz+q^2z^*)\\[0.8ex]
  \varrho_{33}&=\tthird(1+z+z^*)
  \end{array}
\right\}\quad\mbox{or}\quad  
\varrho_{kk}=\tthird+\tfrac{2}{3}\,\mathrm{Re}\bigl(q^{-k}z\bigr)
\,.
\end{equation}
And, as a basic check of consistency, we note that 
$\tr{\varrho_\mathrm{trit}}=1$ is immediate.

Similarly, the diagonal elements $\trho_{mm}$ of 
the Fourier transformed density matrix
$\trho_\mathrm{trit}^{\ }=F\varrho_\mathrm{trit}^{\ }F\adj$
can be expressed in terms of the corresponding complex number
\begin{equation}
  \label{eq:C10}
  Z=\Exp{\I\varphi_1}\varrho_{12}\Exp{-\I\varphi_2}
    +\Exp{\I\varphi_2}\varrho_{23}\Exp{-\I\varphi_3}
    +\Exp{\I\varphi_3}\varrho_{31}\Exp{-\I\varphi_1}\,,
\end{equation}
where the phases $\varphi_j$ are those of (\ref{eq:A4}) and (\ref{eq:A6}).
Explicitly, we have
\begin{equation}
  \label{eq:C11}
\trho_{mm}=\tthird+\tfrac{2}{3}\,\mathrm{Re}\bigl(q^{-m}Z\bigr)  
\end{equation}
as the analog of (\ref{eq:C9}).

\subsection{Pure qutrit states}\label{sec:3paths-pure}
The generic form of the density matrix for a pure qutrit state is
\begin{equation}
  \label{eq:C12}
     \varrho_\mathrm{trit,pure}^{\ }
=\left(\begin{array}{ccc}
p_1 & \sqrt{p_1p_2} & \sqrt{p_1p_3} \\
\sqrt{p_2p_1} & p_2 & \sqrt{p_2p_3} \\
\sqrt{p_3p_1} & \sqrt{p_3p_2} & p_3
\end{array}\right)
=\left(\begin{array}{c}\sqrt{p_1}\\\sqrt{p_2}\\\sqrt{p_3}\end{array}\right)
\bigl(\sqrt{p_1},\sqrt{p_2},\sqrt{p_3}\bigr)
\end{equation}
with $p_1\geq p_2\geq p_3\geq0$ by convention and $p_1+p_2+p_3=1$ by 
normalization.
There are four families of pure states that are of particular importance,
characterized by
\begin{equation}
  \label{eq:C13}
  \begin{array}{l@{\quad}l}
\mbox{Family Ia:} & p_2=p_3\,; \\
\mbox{Family Ib:} & (p_1-p_3)^2+3p_2=1\,; \\
\mbox{Family II:} & p_1=p_2\,; \\
 \mbox{Family III:} & p_3=0\,.
 \end{array}
\end{equation}
Families Ia and Ib have the states of full path knowledge ($p_1=1$,
$p_2=p_3=0$) and of full interference strength ($p_1=p_2=p_3=\tthird$) as
limiting cases. 
The latter is also a member of Family II, whereas the former is in Family III.
The respective other limit of $p_1=p_2=\thalf$, $p_3=0$ is a common member of
Families II and III.
These matters are summarized by the schematic diagram 
\begin{equation}
  \label{eq:C14}
  \raisebox{-70pt}{\includegraphics{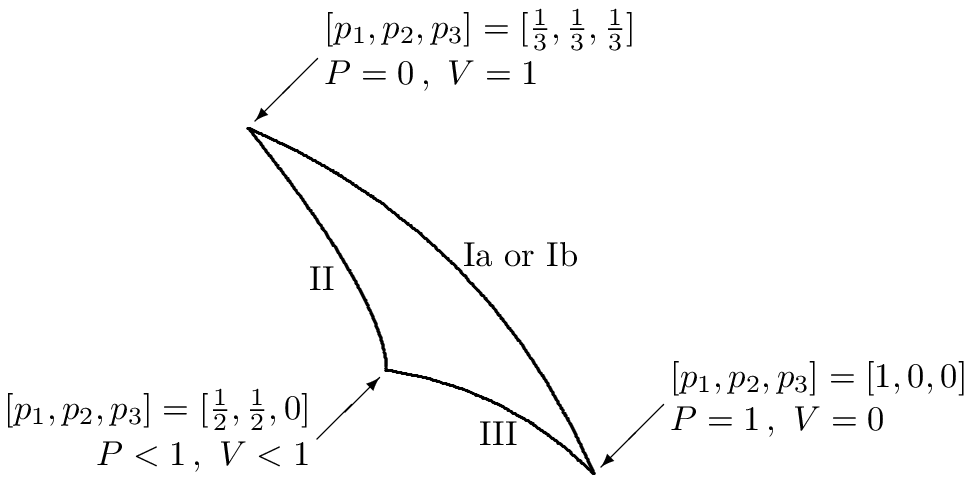}}
\end{equation}
which shows how the families (\ref{eq:C13}) interpolate between the particular
limiting pure states. 

\begin{figure}[!t]
\centering\includegraphics[bb=147 303 477 633]{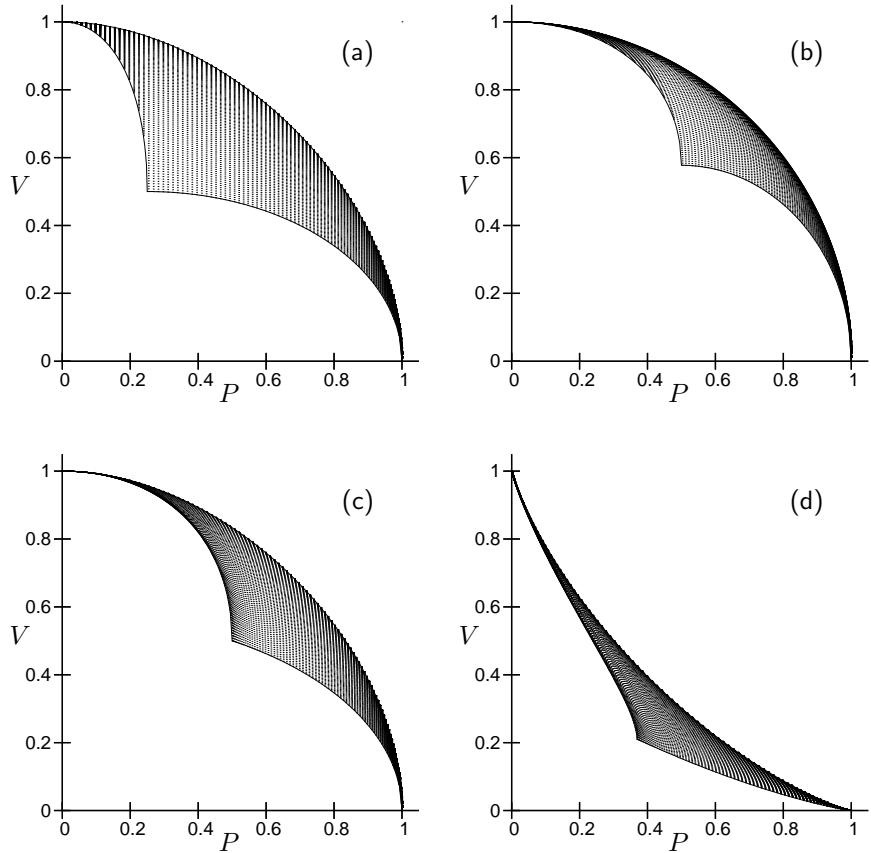}
\caption{\label{fig:qutrit1}%
The $P,V$ values of all pure qutrit states are located in the shaded areas or
on the solid lines enclosing them. 
The four plots refer to quantifying path knowledge (a) by the one-guess bet, 
(b) by the linear bet, (c) by the purity measure, and (d) by the entropic 
measure. 
In all cases, the inner borders are traced out by families II and III of
(\ref{eq:C14}), whereas family Ia resides on the outer border for (a), (c),
and (d) but not for the linear-bet case (b), for which family Ib makes up the
outer border.}   
\end{figure}

In Fig.~\ref{fig:qutrit1}, the families of (\ref{eq:C13}) trace out the borders
in the $P,V$ diagram within which the $P,V$ values of all pure qutrit states
are located.
Despite the obvious differences, all four plots have certain basic features in
common: there is a smooth outer border, and the inner border consists of two
smooth pieces with a cusp where they are joined.
Note in particular that, in marked contrast to the two-path case, not all pure
states give an optimal compromise between $P$ and $V$, only the ones on the
outer border achieve this.  

The cusp for the pure state with $p_1=p_2=\thalf$, $p_3=0$, 
the common state of families II and III, is at 
\begin{equation}
  \label{eq:C15}
(P,V)=\left\{
  \begin{array}{l}
(\tfrac{1}{4},\thalf)\ 
\mbox{for the one-guess bet of Fig.~\ref{fig:qutrit1}a,}\\[1ex]
(\thalf,\tfrac{1}{\sqrt{3}})\ 
\mbox{for the linear-guess bet of Fig.~\ref{fig:qutrit1}b,}\\[1ex]
(\thalf,\thalf)\ 
\mbox{for the purity measure of Fig.~\ref{fig:qutrit1}c,}\\[1ex]
(1-\log_32,\tthird\log_32)\
\mbox{for the entropic measure of Fig.~\ref{fig:qutrit1}d.}
  \end{array}\right.
\end{equation}
The outer border is formed by family Ia, except for the linear-guess bet of
Fig.~\ref{fig:qutrit1}b where the state of family Ib reside on the outer
border.
We now proceed to take a look at the various measures for the path knowledge 
and the implied measures for the interference strength in order to justify
these remarks.

\subsubsection{One-guess bet}\label{sec:3path-1guess}
The path knowledge associated with the one-guess bet, see (\ref{eq:A13}), is
\begin{equation}
  \label{eq:C16}
   P_\mathrm{bet}^{\mathrm{(1\,guess)}}=\tfrac{3}{2}p_1-\thalf\,,
\end{equation}
and the corresponding measure for the interference strength is
\begin{equation}
  \label{eq:C17}
   V_\mathrm{bet}^{\mathrm{(1\,guess)}}=
    \sqrt{p_1p_2}+\sqrt{p_2p_3}+\sqrt{p_3p_1}\,,
\end{equation}
as the maximization required by (\ref{eq:A18}) is an optimization of the phase
factors in (\ref{eq:C10}) which is easily carried out.
We find the borders by maximizing and minimizing
$V_\mathrm{bet}^{\mathrm{(1\,guess)}}$ for a given value of 
$P_\mathrm{bet}^{\mathrm{(1\,guess)}}$, that is:
\begin{equation}
  \label{eq:C18}
  \parbox{0.75\textwidth}{%
      $p_1$ given; 
      $p_2$ in the range $\half(1-p_1)\leq p_2\leq\min\{p_1,1-p_1\}$;
      $p_3=1-p_1-p_2$; then 
      \begin{displaymath}
      \frac{\partial}{\partial p_2} V_\mathrm{bet}^{\mathrm{(1\,guess)}}
        =-\frac{(\sqrt{p_2}-\sqrt{p_3})(\sqrt{p_1}+\sqrt{p_2}+\sqrt{p_3})}
               {2\sqrt{p_2p_3}}\leq0\,,
      \end{displaymath}
     so that $V_\mathrm{bet}^{\mathrm{(1\,guess)}}$ is largest when $p_2$ is
     smallest, and smallest when $p_2$ is largest.}  
\end{equation}
Therefore, we have $p_2=p_3$ on the outer border (family Ia), and $p_1=p_2$ on
the inner border if $p_1\leq\thalf$ (family II) as well as $p_2=1-p_1$ on the
inner border if $p_1\geq\thalf$ (family III). 
In geometrical terms, the outer border is (an arc of) the ellipse
\begin{equation}
  \label{eq:brd-ac}
   2(P+V-\thalf)^2+(P-V)^2=\tfrac{3}{2}\,,
\end{equation}
with the center at $P=V=\tfrac{1}{4}$, the major axis of length $\sqrt{3}$
on the line $V+P=\half$ and the minor axis of length $\sqrt{3/2}$ on the line 
$V=P$.

\subsubsection{Linear bet}\label{sec:3path-lin}
According to (\ref{eq:A14}), we have the path knowledge
\begin{equation}
  \label{eq:C19}
   P_\mathrm{bet}^{\mathrm{(lin)}}=p_1-p_3
\end{equation}
for the linear betting strategy.
Several steps are needed to find the corresponding 
$V_\mathrm{bet}^{\mathrm{(lin)}}$. 
First, we recall the remark after Eq.~(\ref{eq:F2}) and set $\varphi_2=0$ in
\begin{equation}
  \label{eq:C20}
  Z=\sqrt{p_1p_2}\,\Exp{\I\varphi_1}+\sqrt{p_2p_3}\,\Exp{-\I\varphi_3}
    +\sqrt{p_3p_1}\,\Exp{\I(\varphi_3-\varphi_1)}\,.
\end{equation}
Second, we note that the replacements $\varphi_1\to\varphi_1+\frac{2\pi}{3}$,
$\varphi_3\to\varphi_3-\frac{2\pi}{3}$ amount to $Z\to qZ$ and thus permute
the $\trho_{mm}$s of (\ref{eq:C11}) cyclically. 
Therefore, it is permissible to assume that
$\bigl|\trho_{11}-\trho_{22}\bigr|$
is the difference of the largest and the smallest of the $\trho_{mm}$s,
so that
\begin{eqnarray}
  \label{eq:C21}
  V_\mathrm{bet}^{\mathrm{(lin)}}
&=&\frac{2}{\sqrt{3}}\max_{\varphi_1,\varphi_3}\bigl|\mathrm{Im}(Z)\bigr|
\nonumber\\
&=&\frac{2}{\sqrt{3}}\max_{\varphi_1,\varphi_3}
\bigl\{\sqrt{p_1p_2}\,\sin\varphi_1-\sqrt{p_2p_3}\,\sin\varphi_3
    +\sqrt{p_3p_1}\,\sin(\varphi_3-\varphi_1)\bigl\}\,.
\end{eqnarray}
Third, we perform the required maximization over $\varphi_1$ and $\varphi_3$
and arrive at
\begin{eqnarray}
  \label{eq:C22}
&\displaystyle 
   V_\mathrm{bet}^{\mathrm{(lin)}}
   =\frac{2}{\sqrt{3}}\sqrt{p_1p_2+p_2p_3+p_3p_1+2y+3y^2}&
\nonumber\\
&\mbox{where $y\geq0$ solves $2y^3+y^2=p_1p_2p_3$.}&
\end{eqnarray}
If an explicit expression is needed for $y$, then
\begin{equation}
  \label{eq:C23}
  y=\frac{\cos(3\vartheta)}{6\cos\vartheta}\quad\mbox{with}\ 
  \cos(3\vartheta)=\sqrt{27p_1p_2p_3}
\end{equation}
is perhaps the most convenient.

We now search for the largest value of $V_\mathrm{bet}^{\mathrm{(lin)}}$
for a given value of $P_\mathrm{bet}^{\mathrm{(lin)}}$, that is the difference
$P\equiv p_1-p_3$ is fixed.
The permissible values of $p_2$ are then in the range
\begin{equation}
  \label{eq:C24}
  \tthird(1-P)\leq p_2\leq\left\{
    \begin{array}{cl}
 \tthird(1+P) &\mbox{for $P\leq\half$,}\\[1ex]
1-P &\mbox{for $P\geq\half$.}
    \end{array}\right.
\end{equation}
If $p_2$ equals its lower bound, the state is in family Ia; at the upper
bounds we have family II or III, respectively.

We note that $\D p_1=\D p_3=-\half\D p_2$, with the consequence
\begin{equation}
  \label{eq:C25}
  \frac{\D}{\D p_2}{V_\mathrm{bet}^{\mathrm{(lin)}}}^2=
   \frac{1}{3y}\bigl[(1-3p_2)(1-p_2+2y)-P^2\bigr]\,.
\end{equation}
At the bounds on $p_2$ in (\ref{eq:C24}) we thus have
\begin{eqnarray}
  \label{eq:C26a}
  &&\frac{\D}{\D p_2}{V_\mathrm{bet}^{\mathrm{(lin)}}}>0
\mbox{\ at the lower bound,}\nonumber\\
&\mbox{and}&\frac{\D}{\D p_2}{V_\mathrm{bet}^{\mathrm{(lin)}}}<0
\mbox{\ at the upper bounds.}
\end{eqnarray}
Therefore,  $V_\mathrm{bet}^{\mathrm{(lin)}}$ has local minima for families
Ia, II, and III. 
The smaller value is always obtained for the upper bounds in (\ref{eq:C24}),
as one can verify after first observing that
\begin{equation}
  \label{eq:C26b}
    \begin{array}{r@{\quad}l}
\mbox{lower bound:} & p_1=\tthird(1+2P)\,,\ p_2=p_3=\tthird(1-P)\,,
\\ & 2y^2+p_1y=p_1p_3\,; \\[1ex]
\mbox{upper bound for $P\leq\half$:} & 
p_1=p_2=\tthird(1+P)\,,\ p_3=\tthird(1-2P)\,,\\
& 2y^2+p_3y=p_1p_3\,; \\[1ex]
 \mbox{upper bound for $P\geq\half$:} & p_1=P\,,\ p_2=1-P\,,\ p_3=0\,,\ y=0\,;
 \end{array}
\end{equation}
so that families II and III trace out the inner borders, indeed. 

Further, Eq.~(\ref{eq:C25}) implies that 
$V_\mathrm{bet}^{\mathrm{(lin)}}$ is maximal at the intermediate
$p_2$ value that obeys
\begin{equation}
  \label{eq:C27}
  (1-3p_2)(1-p_2+2y)=P^2
\end{equation}
with $y$ from (\ref{eq:C22}). 
Therefore, the pure state with
\begin{eqnarray}
  \label{eq:C28}
 &&p_1=\tfrac{1}{6}(1+P)(2+P)\,,\quad p_2=\tthird(1-P^2)\,,\nonumber\\
 && p_3=\tfrac{1}{6}(1-P)(2-P)\,,\quad y=\tfrac{1}{6}(1-P^2)\,,
\end{eqnarray}
maximizes $V_\mathrm{bet}^{\mathrm{(lin)}}$ for given  
$P=P_\mathrm{bet}^{\mathrm{(lin)}}$.  
Indeed, the outer border of Fig.~\ref{fig:qutrit1}b is traced by family Ib of
(\ref{eq:C13}). 
And since $V_\mathrm{bet}^{\mathrm{(lin)}}=\sqrt{1-P^2}$ for the pure states
specified by (\ref{eq:C28}), the outer border is the quarter circle
\begin{equation}
  \label{eq:brd-b}
  P^2+V^2=1\,.
\end{equation}

\subsubsection{Purity}\label{sec:3path-pur}
It is a matter of inspection to verify that the path-knowledge function of
Sec.~\ref{sec:P-Durr} is given by the modulus of $z$,
\begin{equation}
  \label{eq:C29}
  P_\mathrm{pur}=|z|\,.
\end{equation}
It follows that the induced interference-strength measure is
\begin{equation}
  \label{eq:C30}
   V_\mathrm{pur}=\max_{\varphi_j}\bigl\{|Z|\bigr\}
=\left|\varrho_{12}\phstar\right|+\left|\varrho_{23}\phstar\right|
+\left|\varrho_{31}\phstar\right|\,,
\end{equation}
because the maximal value of $|Z|$ obtains when the phases $\varphi_j$ are
just the ones that exhibit the common phase factor $\exp(\third\I\theta)$ of 
(\ref{eq:C4}).
These equations apply to pure or mixed states. 

We note that, although $P_\mathrm{pur}$ is D\"urr's path knowledge function of
Ref.~\refcite{Durr:01}, the corresponding interference strength
$V_\mathrm{pur}$ of (\ref{eq:C30}) is \emph{not} the one suggested by D\"urr,
which is
\begin{equation}
  \label{eq:C31}
   \overline{V}_\mathrm{pur}=\sqrt{3\bigl(\left|\varrho_{12}\phstar\right|^2
+\left|\varrho_{23}\phstar\right|^2
+\left|\varrho_{31}\phstar\right|^2\bigr)}\,.
\end{equation}
Therefore, D\"urr's pair $P_\mathrm{pur},\overline{V}_\mathrm{pur}$ does not
fit into the general strategy of Sec.~\ref{sec:General}.
Rather than linking $V$ to $P$ by (\ref{eq:A18}), his choice is such that
\begin{equation}
  \label{eq:C32}
  P^2+V^2=\tfrac{3}{2}\tr{\varrho_\mathrm{trit}^2}-\thalf\leq1
\end{equation}
by construction.
The equal sign holds for \emph{all} pure states, as it does in 
Eq.~(\ref{eq:P+V-bit}).

For pure states, Eqs.~(\ref{eq:C29}) and (\ref{eq:C30}) give
\begin{eqnarray}
  \label{eq:C33}
  P_\mathrm{pur}&=&\sqrt{1-3(p_1p_2+p_2p_3+p_3p_1)}\,,
\nonumber\\
  V_\mathrm{pur}&=&\sqrt{p_1p_2}+\sqrt{p_2p_3}+\sqrt{p_3p_1}\,.
\end{eqnarray}
We note the coincidence that 
$P_\mathrm{pur}=P_\mathrm{bet}^{\mathrm{(1\,guess)}}$ 
and $V_\mathrm{pur}=V_\mathrm{bet}^{\mathrm{(1\,guess)}}$ on the outer border,
traced out by family Ia,
so that the outer border for the purity measures is also 
the ellipse (\ref{eq:brd-ac}) of the one-guess bet.

\subsubsection{Entropy}\label{sec:3path-ent}
For the pure qutrit states of (\ref{eq:C12}), the entropic measures for path
knowledge and interference strength are
\begin{equation}
  \label{eq:C34}
  P_\mathrm{ent}
  =\frac{1}{\log3}\bigl[p_1\log(3p_1)+p_2\log(3p_2)+p_3\log(3p_3)\bigr]
\end{equation}
and
\begin{equation}
  \label{eq:C35}
  V_\mathrm{ent}
  =\frac{1}{\log3}\bigl[\tp_1\log(3\tp_1)+\tp_2\log(3\tp_2)
                        +\tp_3\log(3\tp_3)\bigr]
\end{equation}
with
\begin{eqnarray}
  \label{eq:C36}
  \tp_1&=&\tthird
        +\tfrac{2}{3}\bigl(\sqrt{p_1p_2}+\sqrt{p_2p_3}+\sqrt{p_3p_1}\bigr)\,,
\nonumber\\
  \tp_2=\tp_3&=&\tthird
        -\tthird\bigl(\sqrt{p_1p_2}+\sqrt{p_2p_3}+\sqrt{p_3p_1}\bigr)\,.
\end{eqnarray}
It turns out that, here too, the states of families II and III form the inner
borders while the states of family Ia make up the outer border once more.
Accordingly, 
\begin{eqnarray}
  \label{eq:brd-d}
  P_\mathrm{ent}
    &=&\frac{1}{3\log3}\bigl[(1+2u)\log(1+2u)+2(1-u)\log(1-u)\bigr]\,,
\nonumber\\
  V_\mathrm{ent}
    &=&\frac{1}{3\log3}\bigl[(1+2v)\log(1+2v)+2(1-v)\log(1-v)\bigr]\,,
\nonumber\\
&&\mbox{with $(u-v)^2+2(u+v-\thalf)^2=\tfrac{3}{2}$ for $0\leq u,v\leq1$}
\end{eqnarray}
is a convenient parameterization of the outer border in 
Fig.~\ref{fig:qutrit1}d. 
Note that the parameters $u,v$ have values on the ellipse of
(\ref{eq:brd-ac}).

\begin{figure}[!t]
\centering\includegraphics{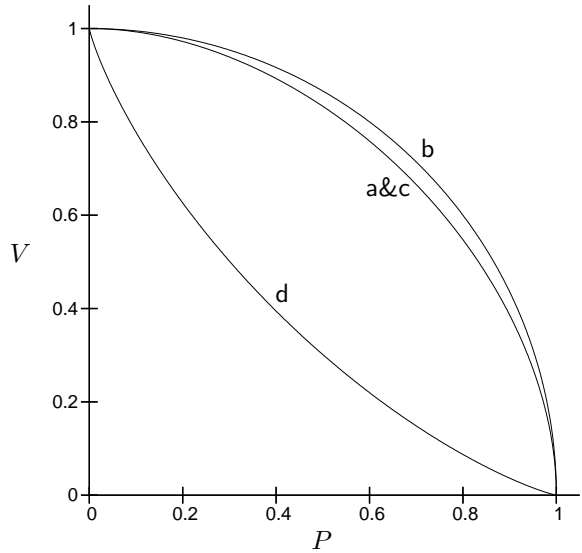}
\caption{\label{fig:qutrit2}%
The outer borders of the shaded areas in Fig.~\ref{fig:qutrit1}.
For Figs.~\ref{fig:qutrit1}a and \ref{fig:qutrit1}c we have
curve~\textsf{a\&c}, the ellipse of Eq.~(\ref{eq:brd-ac}).
Curve~\textsf{b} is the circle of Eq.~(\ref{eq:brd-b}), the outer border in  
Fig.~\ref{fig:qutrit1}b.
For the entropic measures of Fig.~\ref{fig:qutrit1}d, the $P,V$ values of
(\ref{eq:brd-d}) are shown as curve~\textsf{d}.}
\end{figure}

\subsection{Outer borders}\label{sec:trit-2of3}
The outer borders found for the four cases that are examined in 
Secs.~\ref{sec:3path-1guess}--\ref{sec:3path-ent} are shown in
Fig.~\ref{fig:qutrit2}.
The possible values for $P$ and $V$ are restricted to the area bounded by the
respective outer border and the axes.
Of the four choices, the smallest area is that for the entropic quantities,
and the largest area is the quarter-circle of the linear bet.

But this is not the absolutely largest area. 
It is indeed possible to have permissible $P,V$ pairs almost everywhere inside
the square $0\leq P,V\leq1$.
To demonstrate this point, we consider a general bet, for which
\begin{equation}
  \label{eq:C37}
  P_\mathrm{bet}=p_1+g_2p_2-(1+g_2)p_3\quad\mbox{with $1>g_2\geq-\half\,,$}
\end{equation}
and the pure qutrit state specified by
\begin{equation}
  \label{eq:C38}
  \varrho_\mathrm{trit}^{\ }=\half\left(
    \begin{array}{rr@{\quad}r}1&-1&0\\-1&1&0\\0&0&0 \end{array}\right)\,,
\end{equation}
so that $p_1=p_2=\half$, $p_3=0$ and
\begin{equation}
  \label{eq:C39}
  P_\mathrm{bet}=\half(1+g_2)\,.
\end{equation}
For either one of the two $3\times3$ Fourier matrices in (\ref{eq:F4}) we have
$F\varrho_\mathrm{trit}^{\ }F\adj=\varrho_\mathrm{trit}^{\ }$, implying
\begin{equation}
  \label{eq:C40}
  \half(1+g_2)\leq V_\mathrm{bet}<1\,.
\end{equation}
By now choosing $g_2$ sufficiently close to, but less than, unity, 
we can push $(P_\mathrm{bet},V_\mathrm{bet})$ as close to the corner
$(P,V)=(1,1)$ as we wish.
As a consequence of criterion (\ref{eq:A8}a), 
the limiting value $g_2=1$ is not permitted in (\ref{eq:A10}); if it were, it
would realize a bet on two of the three paths versus the third. 
In practice, however, a $g_2$ value that is rather close to $1$ will implement
such a 2-of-3 bet quite well, and then the two dominating paths can interfere
with almost full strength even when almost perfect 2-of-3 path knowledge is at
hand.

For the pure qutrit states of (\ref{eq:C12}) we have
\begin{eqnarray}
  \label{eq:C41}
   P_\mathrm{bet}&\to&1-3p_3\nonumber\\
\quad\mbox{and}\enskip
   V_\mathrm{bet}&\to&\left\{\begin{array}{cl}
     1 & \mbox{if $\sqrt{p_1}\leq\sqrt{p_2}+\sqrt{p_3}$\,,}\\[1ex]
2\bigl(\sqrt{p_1p_2}-\sqrt{p_2p_3}+\sqrt{p_3p_1}\bigr) 
       & \mbox{if $\sqrt{p_1}\geq\sqrt{p_2}+\sqrt{p_3}$\,,}
     \end{array}\right.
\end{eqnarray}
in the limit $g_2\to1$.
Clearly, all states with $p_1=p_2$, those of family II in  (\ref{eq:C13}),
have $V_\mathrm{bet}=1$ in this limit, so that the outer border is given by
\begin{equation}
  \label{eq:C42}
  \max\{P,V\}=1\,,
\end{equation}
and the whole area of the square $0\leq P,V\leq1$ is covered.

\section{Four-path interferometers: Ququarts}\label{sec:Quarts}
In an interferometer with four paths, the interfering object is a quaternary
quantum alternative or \emph{ququart}.%
\footnote{We follow the practice of recent publication, 
such as Ref.~\refcite{Kulik+3:06},
although \emph{ququad} could be preferable, inasmuch as 
$1\,\mathrm{quad}=2\,\mathrm{bits}$ according to the 
Hackers' Dictionary,\cite{HackDict} which offers \emph{tayste} as a serious
alternative to \emph{quad} and \emph{crumb} as a silly alternative but does
not mention \emph{quart}.  
Footnote \ref{fn:4} applies \textit{mutatis mutandis}.}
As a consequence of the parameter $t$ in the $4\times4$ Fourier matrix
(\ref{eq:F5}) and the nonequivalence of column permutations and row
permutations, the analysis of four-path interferometers is much more involved
than that of interferometers with two or three paths and has not been carried
out in full as yet.
We shall therefore be brief and only discuss two issues: 
Two conjectures, one disproven by a counter example and the other undecided, 
and a recent analysis of a four-path interferometer with the tools of 
two-path interferometry.

\subsection{Linear bet: One conjecture rejected, another proposed}%
\label{sec:4path-lin}
Whereas the border lines in the $P,V$-diagram are different, as a rule, for
two-path and three-path interferometers, it is remarkable that we have the
same quarter-circle border for the linear bet in Figs.~\ref{fig:qubit} and
\ref{fig:qutrit2}. 
It is tempting to conjecture that the linear bet values of $(P,V)$ are in the
quarter-circle area $P^2+V^2\leq1$ also in the case of four-path
interferometers, or perhaps even quite generally for $n$-path interferometers.
This conjecture is, however, false as can be demonstrated by the following
counter example.\cite{WeiHuiThesis}

Consider the pure-case $4\times4$ density matrix
\begin{equation}
  \label{eq:D1}
  \varrho=\left(\begin{array}{c}%
\sqrt{p_4}\\\sqrt{p_3}\\\sqrt{p_2}\\\sqrt{p_1}%
\end{array}\right)
\bigl(\sqrt{p_4},\sqrt{p_3},\sqrt{p_2},\sqrt{p_1}\bigr)
\end{equation}
with the path probabilities
\begin{equation}
  \label{eq:D2}
  p_1=\frac{1}{10}G^{-4}\,,\quad
  p_2=\frac{1}{10}G^{-2}\,,\quad
  p_3=\frac{1}{10}G^{2}\,,\quad
  p_4=\frac{1}{10}G^{4}\,,
\end{equation}
where $G=\frac{1}{2}(\sqrt{5}-1)$ is the golden ratio.
The $t=0$ Fourier matrix (\ref{eq:F5}) yields $\trho=\varrho$, so that 
$V_\mathrm{bet}^{\mathrm{(lin)}}(\varrho)\geq %
P_\mathrm{bet}^{\mathrm{(lin)}}(\trho)=P_\mathrm{bet}^{\mathrm{(lin)}}(\varrho)%
=\sqrt{5/9}$, and
\begin{equation}
  \label{eq:D3}
{P_\mathrm{bet}^{\mathrm{(lin)}}}^2+{V_\mathrm{bet}^{\mathrm{(lin)}}}^2%
\geq\frac{10}{9}>1  
\end{equation}
for this $\varrho$.

\begin{figure}[!t]
\centering\includegraphics{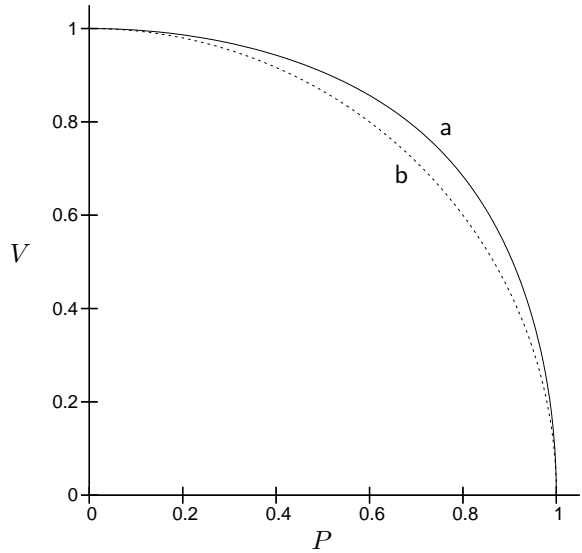}
\caption{\label{fig:ququart}%
Possible values for the path-knowledge measure $P$ and the
interference-strength measure $V$ in four-path interferometers.
Curve \textsf{a} is the conjectured border line for 
$(P,V)=(P_\mathrm{bet}^\mathrm{(lin)},V_\mathrm{bet}^\mathrm{(lin)})$
that is traced out by (\ref{eq:D6}),
and curve \textsf{b} is the quarter circle $P^2+V^2=1\,$.}
\end{figure}

There is numerical evidence, 
but no clear-cut demonstration of the case as yet, 
that the linear-bet border is traced out by pure states with density
matrices of the form (\ref{eq:D1}) with
\begin{eqnarray}
  \label{eq:D4}
 &&\sqrt{p_1/p_2}=\sqrt{p_3/p_4}=\Exp{\vartheta}\,,\quad
  \sqrt{p_3/p_1}=\sqrt{p_4/p_2}=\tanh\frac{\theta}{2}\nonumber\\
&&\mbox{where}\quad 4\sinh\vartheta\,\sinh\theta=1\,,
\end{eqnarray}
and the optimal Fourier transform being the $t=1$ version of (\ref{eq:F5}).
Then we have
\begin{equation}
  \label{eq:D5}
 \sqrt{\tp_1/\tp_2}=\sqrt{\tp_3/\tp_4}=\Exp{\theta}\,,\quad
  \sqrt{\tp_3/\tp_1}=\sqrt{\tp_4/\tp_2}=\tanh\frac{\vartheta}{2}\,,  
\end{equation}
which are reciprocal to the relations in (\ref{eq:D4}), and
\begin{equation}
  \label{eq:D6}
  P_\mathrm{bet}^{\mathrm{(lin)}}
  =\frac{1}{3}\tanh\vartheta+\frac{2}{3}\sech\theta\,,\quad
  V_\mathrm{bet}^{\mathrm{(lin)}}
  =\frac{1}{3}\tanh\theta+\frac{2}{3}\sech\vartheta
\end{equation}
parameterize the thus conjectured border line.
Figure \ref{fig:ququart} shows that, except for $(P,V)=(1,0)$ and
$(P,V)=(0,1)$, all values of (\ref{eq:D6}) have 
${P_\mathrm{bet}^{\mathrm{(lin)}}}^2+{V_\mathrm{bet}^{\mathrm{(lin)}}}^2>1$.
The values of (\ref{eq:D2}) and (\ref{eq:D3}) obtain for 
$\sinh\vartheta=\sinh\theta=\half$.

\subsection{$4\neq2\times2$}\label{sec:4path-Gershon}
In a recent series of papers, there is a claim that one can violate the
duality relation (\ref{eq:P+V-bit}), thereby ``cheating on complementarity''
in this manner; see Ref.~\refcite{Kurizki+4:06}, for instance.
Of course, Bohr's Principle of Complementarity,%
\footnote{See Ref.~\refcite{LQM-PP} for a textbook discussion.}
the fundamental principle of quantum kinematics, remains untouched: 
in fact, the authors of Ref.~\refcite{Kurizki+4:06} are relying on it in their
analysis, as all arguments do that invoke the quantum formalism.

More importantly, the thought experiment in question --- a Mach-Zehnder (MZ) 
set-up traversed by a qubit, so that the spatial binary alternative of the
MZ geometry and the internal qubit together make up a ququart ---
is a four-path interferometer%
\footnote{In the MZ set-up of Ref.~\refcite{Kurizki+4:06} there are fixed
relations between the phases $\phi_1,\dots,\phi_4$ of the four path
amplitudes, so that the full flexibility needed in (\ref{eq:A18}) would
require a modification of the set-up.}
to which the two-path relation (\ref{eq:P+V-bit}) does not apply.
When applying it nevertheless, by letting path knowledge refer solely to the
arms of the MZ set-up, the actual paths of the four-path
interferometer are grouped into two pairs, and we have a 2-of-4 bet,
similar to the 2-of-3 bet discussed in Sec.~\ref{sec:trit-2of3}.

The situation is then that of a limiting four-path bet with $g_1=1$, $g_4=-1$,
and $1>g_2=-g_3\to1$, and it is easy to have both $P$ and $V$ close to unity.
But this has no bearing on (\ref{eq:P+V-bit}) or any other relation for
two-path interferometers, and poses no challenge to complementarity.

An analysis of the ququart MZ set-up as a four-path interferometer with the
methodology introduced in Sec.~\ref{sec:General} would be of great interest.
If the four paths are implemented by having polarized photons in the two arms
of a MZ interferometer, the particle mode and the wave mode of
Fig.~\ref{fig:nPaths} can be realized by known and rather simple 
tools.\cite{Englert+2:01}

\section{$n$-path interferometers: Qunits}\label{sec:Nits}
In a multi-path interferometer with $n$ paths, the generic situation of
Fig.~\ref{fig:nPaths}, the interfering object is a $n$-fold quantum
alternative, or \emph{qunit}.%
\footnote{Other authors prefer to speak of \emph{qudits}, having
$d$-dimensional Hilbert subspaces in mind.
Here too, footnote \ref{fn:4} applies \textit{mutatis mutandis}.}
Little is known about the possible $(P,V)$ values for qunits.
We can only offer some observations about path knowledge measures that involve
solely $p_1$ and put on record
an unproven conjecture about the pair $(P_\mathrm{ent},V_\mathrm{ent})$ of
entropic measures that is suggested by a pattern observed for
qubits and qutrits, and by some numerical evidence.

\begin{figure}[!t]
\centering\includegraphics{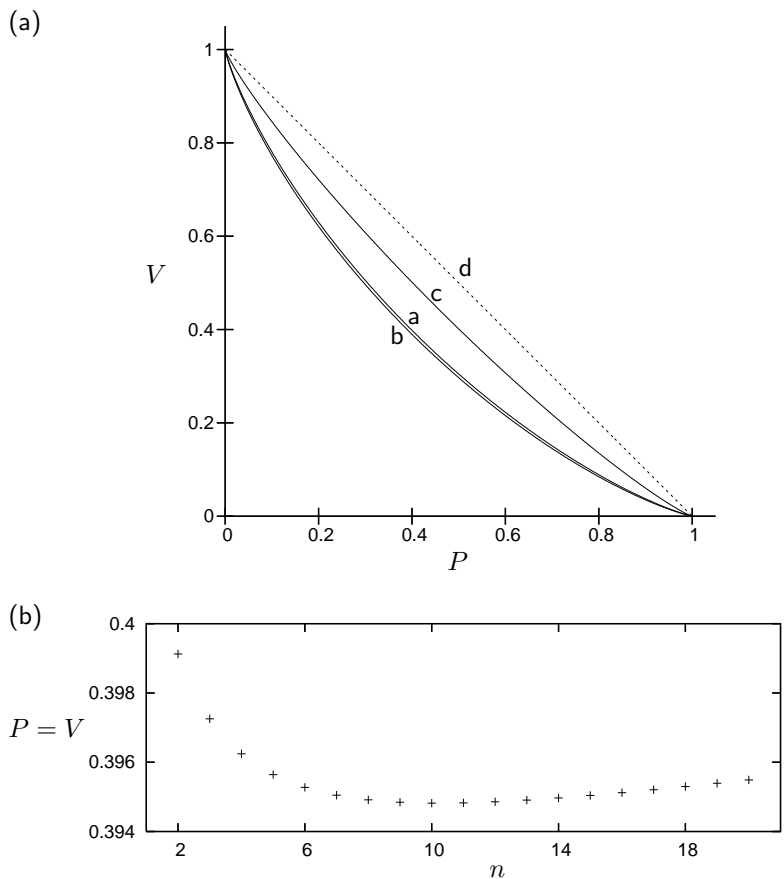}
\caption{\label{fig:qunit}%
Conjectured border lines for the entropic measures for path knowledge and
interference strength in multi-path interferometers.
The top plot (a) shows the border lines associated with (\ref{eq:E1})
and (\ref{eq:E2}), whereby curve \textsf{a} is for $n=2$, curve \textsf{b} for
$n=10$, curve \textsf{c} for $n=10^7$, and curve \textsf{d} for $n\to\infty$. 
---
The bottom plot (b) shows the symmetric $P=V$ values for
$n=2,3,\dots,20$ and confirms that the smallest value is found for $n=10$.
}
\end{figure}

\subsection{Entropic measures: A conjecture}
The conjecture is this: The border lines for the entropic measures for path
knowledge and interference strength are traced out by the pure states whose
$n\times n$ density matrices are of the form
\begin{equation}
  \label{eq:E1}
  \varrho=\left(\begin{array}{c}%
\sqrt{p_2}\\\sqrt{p_2}\\\vdots\\\sqrt{p_2}\\\sqrt{p_1}%
\end{array}\right)
\bigl(\sqrt{p_2},\sqrt{p_2},\ldots,\sqrt{p_2},\sqrt{p_1}\bigr)
\qquad\mbox{with}\quad p_1+(n-1)p_2=1\,,
\end{equation}
and the optimal Fourier matrix is the standard one of
(\ref{eq:A4})---(\ref{eq:A6}) with $\varphi_j=0$.
Then $\trho$ has the same form with $p_1$ and $p_2$ replaced by
\begin{equation}
  \label{eq:E2}
  \tp_1=\frac{1}{n}\bigl(\sqrt{p_1}+(n-1)\sqrt{p_2}\bigr)^2
\qquad\mbox{and}\qquad
  \tp_2=\frac{1}{n}\bigl(\sqrt{p_1}-\sqrt{p_2}\bigr)^2\,,
\end{equation}
respectively.
The identities
\begin{equation}
  \label{eq:E3}
  np_1+n\tp_1-2\sqrt{np_1\tp_1}=n-1\,,\quad np_2+n\tp_2+2\sqrt{np_2\tp_2}=1
\end{equation}
exhibit the reciprocal nature of the mappings
$(p_1,p_2)\leftrightarrow(\tp_1,\tp_2)$.  

Thus surmising that these pure states are on the border of the
$(P_\mathrm{ent},V_\mathrm{ent})$ values in the $P,V$-diagram, we get the
border lines of Fig.~\ref{fig:qunit}(a). 
For moderate $n$ values, the border lines are very similar to the $n=2$ line
in Fig.~\ref{fig:qubit} and the $n=3$ line in Fig.~\ref{fig:qutrit2}, whereas
we get the straight line $P_\mathrm{ent}+V_\mathrm{ent}=1$ in the limit of
$n\to\infty$. 

The conjectured border lines of Fig.~\ref{fig:qunit}(a) have the remarkable
feature that the area of permissible $(P_\mathrm{ent},V_\mathrm{ent})$ values
decreases from $n=2$ to $n=10$ and then increases.%
\footnote{We resist the strong temptation to suggest that this particular role
that $n=10$ plays in the context of wave-particle duality explains why humans
have ten fingers.}
This is illustrated by the plot in Fig.~\ref{fig:qunit}(b) which shows, for
$n=2,3,\ldots,20$, the symmetric values 
\begin{equation}
  \label{eq:E4}
   P_\mathrm{ent}=V_\mathrm{ent}=\log_n\frac{\sqrt{n}}{2}
                               +\frac{1}{\sqrt{n}}\log_n(\sqrt{n}+1)
\end{equation}
that one gets for
\begin{equation}
  \label{eq:E5}
  p_1=\tp_1=\frac{\sqrt{n}+1}{2\sqrt{n}}\,,\qquad
  p_2=\tp_2=\frac{1}{2\sqrt{n}(\sqrt{n}+1)}\,,
\end{equation}
with the right-hand side of (\ref{eq:E4}) equal to $0.394845$, $0.394820$, and
$0.394827$ for $n=9$, $10$, and $11$, respectively.
We leave it at that.

\subsection{Measures based solely on %
$\displaystyle p_1=\max_m\{\varrho_{mm}\}$}
In (\ref{eq:A13}),  (\ref{eq:A17c}), and (\ref{eq:A17c}) we encounter path
knowledge measures that involve only 
$\displaystyle p_1=\max_m\{\varrho_{mm}\}$: the measure of the one-bet guess
$P_\mathrm{bet}^{\mathrm{(1\,guess)}}$ and the extreme limits
$P_\mathrm{Ren}^{(\infty)}$ and $P_\mathrm{Ren}^{(0)}$ of the R\'enyi-type
measures.  
For these, it is clear that the pure states of (\ref{eq:E1})--(\ref{eq:E3})
maximize $V$ for given $P$, and $P$ for given $V$.

Another way of writing the first identity in (\ref{eq:E3}),
\begin{equation}
  \label{eq:E6}
  n\bigl(p_1+\tp_1-1\bigr)^2+\frac{n}{n-1}\bigl(p_1-\tp_1)^2=1\,,
\end{equation}
gives us the border lines for $P_\mathrm{bet}^{\mathrm{(1\,guess)}}$ and
$P_\mathrm{Ren}^{(\infty)}$.
In the case of the one-guess bet, we have $P=(np_1-1)/(n-1)$ 
and $V=(n\tp_1-1)/(n-1)$, and it follows that the border line
is an arc of the ellipse
\begin{equation}
  \label{eq:E7}
  \frac{(n-1)^2}{n}\Bigl(P+V-\frac{n-2}{n-1}\Bigr)^2
                      +\frac{n-1}{n}\bigl(P-V\bigr)^2=1\,.
\end{equation}
As it should, this is the ellipse of (\ref{eq:brd-ac}) for $n=3$, and the
circle $P^2+V^2=1$ for $n=2$.
The ellipse of (\ref{eq:E7}) is centered at 
$P=V=(n-2)/(2n-2)$; its major axis of length 
$\sqrt{2n/(n-1)}$ is on the line 
$V+P=(n-2)/(n-1)$, and the minor axis of length
$\sqrt{2n}/(n-1)$ is on the line $P=V$. 

In case of the R\'enyi-type measure for $\lambda\to\infty$, it is $P=p_1$ and
$V=\tp_1$ except when $p_1=1,\tp_1=1/n$ or $p_1=1/n,\tp_1=1$,
and we have the arc with $P>1/n$ and $V>1/n$ of the ellipse
\begin{equation}
  \label{eq:8}
    n\bigl(P+V-1\bigr)^2+\frac{n}{n-1}\bigl(P-V)^2=1
\end{equation}
as part of the border line. 
The straight lines with $P\leq1/n$ and $V=1$ or $P=1$ and
$V\leq1/n$ complete the border line, but the points on these straight
segments are not permitted, except for the corners at $(P,V)=(1,0)$ and
$(P,V)=(0,1)$. 

In the limit $n\to\infty$, the straight line $P+V=1$ is the border for 
$(P_\mathrm{bet}^{\mathrm{(1\,guess)}},V_\mathrm{bet}^{\mathrm{(1\,guess)}})$
and $(P_\mathrm{Ren}^{(\infty)},V_\mathrm{Ren}^{(\infty)})$.
The same line is conjectured above for $(P_\mathrm{ent},V_\mathrm{ent})$ for
$n\to\infty$. 

By contrast, matters are quite simple in the $\lambda\to0$ limit of the
R\'enyi-type measures.
All permissible values for $(P_\mathrm{Ren}^{(0)},V_\mathrm{Ren}^{(0)})$ have
either $P_\mathrm{Ren}^{(0)}=0$ or $V_\mathrm{Ren}^{(0)}=0$.

\section{Summary}
We have presented a systematic way of quantifying path knowledge and
interference strength in multi-path interferometers. 
The quantitative measures for particle aspects (path knowledge) and wave
aspects (interference strength) have a clear operational meaning and are
naturally linked to each other. 
This systematic link, which exploits general Fourier transformations,
distinguishes our approach from earlier attempts.

Since there is no unique procedure for assigning a single number to the path
knowledge when there are more than two paths, we have discussed several
self-suggesting definitions of the path-knowledge measure $P$ and the induced
interference-strength measure $V$.  
As a consequence of wave-particle duality, $P=1$ implies $V=0$ and $V=1$
implies $P=0$ for all choices, but the range of values allowed for the pair
$(P,V)$ depends on the particular choice.

We have illustrated our approach with the familiar example of two-path
interferometers and a thorough analysis of three-path interferometers, and
have given glimpses at four-path interferometers and general multi-path
interferometers. 
This sets the stage for further studies.

Perhaps the most important step to be taken now is an investigation
the multi-path analog 
of the transition from \emph{predictability} to \emph{distinguishability} 
in two-path interferometers.\cite{Englert:96}
We will report from this front in due course.

\section*{Acknowledgments}
BGE is grateful for the hospitality extended to him by Christian Miniatura
at the Institut Nonlineaire de Nice in 2005 and 2007, where part of this work
was carried out.
We acknowledge support by A*STAR Temasek Grant 012-104-0040 and by 
NUS Grant R-144-000-179-112. 
Centre for Quantum Technologies is a Research Centre of Excellence funded by
Ministry of Education and National Research Foundation of Singapore.


\begin{thebibliography}{99}

\bibitem{Einstein:05}
A. Einstein,
\JournalTitle{Ann.\ Physik}\textbf{17}, 132 (1905);
English translation in Ref.~\refcite{terHaar:67}.

\bibitem{Bohr:28}
N. Bohr,
\JournalTitle{Die Naturwissenschaften}\textbf{16}, 245 (1928);
English Version: \JournalTitle{Nature}\textbf{121}, 580 (1928);
the latter is reprinted in Ref.~\refcite{Wheeler+1:83}.

\bibitem{Bohr:49}
N. Bohr, ``Discussions with Einstein on Epistemological Problems in Atomic
Physics'' in \textit{Albert Einstein: Philosopher--Scientist},
ed.\ P.~A.~Schilpp (Library of Living Philosophers, Evanston, 1949);
reprinted in Ref.~\refcite{Wheeler+1:83}. 

\bibitem{Held:98}
C. Held, \textit{Die Bohr-Einstein-Debatte\/}
(Schoeningh, Paderborn, 1998).

\bibitem{Englert+1:00}
B.-G. Englert and J. A. Bergou,
\OC\textbf{179}, 337 (2000). 

\bibitem{Wootters+1:79}
W. K. Wootters and W. H. Zurek,
\PRD\textbf{19}, 473 (1979).

\bibitem{Rauch+1:84}
H. Rauch and J. Summhammer,  
\PL\textbf{A104}, 44 (1984).

\bibitem{Glauber:86}
R. Glauber, 
\JournalTitle{Ann.\ N. Y. Acad.\ Sci.}\textbf{480}, 336 (1986).

\bibitem{Mittelstaedt+2:87}
P. Mittelstaedt, A. Prieur, and R. Schieder, 
\FP\textbf{17}, 891 (1987).

\bibitem{Greenberger+1:88}
D. M. Greenberger and A. Yasin,  
\PL\textbf{A128}, 391 (1988).

\bibitem{Mandel:91}
L. Mandel, 
\JournalTitle{Opt.\ Lett.}\textbf{16}, 1882 (1991).

\bibitem{Jaeger+2:95}
G. Jaeger, A. Shimony, and L. Vaidman, 
\PRA\textbf{51}, 54 (1995).

\bibitem{Englert:96}
B.-G. Englert, 
\PRL\textbf{77}, 2154 (1996).

\bibitem{Englert:99}
B.-G. Englert, 
\JournalTitle{Z.\ Naturforsch.}\textbf{54a}, 11 (1999).

\bibitem{Baldzuhn+2:89}
J. Baldzuhn, E. Mohler, and W. Martienssen,  
\ZPhys\textbf{B77}, 347 (1989).

\bibitem{Baldzuhn+1:91}
J. Baldzuhn and W. Martienssen,  
\ZPhys\textbf{B82}, 309 (1991).

\bibitem{Schwindt+2:99}
P. D. D. Schwindt, P. G. Kwiat, and B.-G. Englert,
\PRA\textbf{60}, 4285 (1999).

\bibitem{Kwiat+2:99}
P. G. Kwiat, P. D. D. Schwindt, and B.-G. Englert,
``What does a quantum eraser erase?''
in \textit{Mysteries, Puzzles, and Paradoxes in Quantum Mechanics},
ed.\ R. Bonifacio (CP461, The American Institute of Physics, 1999).

\bibitem{Trifonov+3:02}
A. Trifonov, G. Bj\"ork, J. S\"oderholm,  and T. Tsegaye, 
\JournalTitle{Eur.\ Phys.\ J.\ D}\textbf{18}, 251 (2002).

\bibitem{Walborn+3:02}
S. P. Walborn, M. O. Terra Cunha, S. P\'adua, and C. H. Monken,
\PRA\textbf{65}, 033818 (2002).

\bibitem{Summhammer+2:87}
J. Summhammer, H. Rauch, and D. Tuppinger,
\PRA\textbf{36}, 4447 (1987).

\bibitem{Durr+2:98a}
S. D\"{u}rr, T. Nonn,  and G. Rempe, 
\JournalTitle{Nature}\textbf{395}, 33 (1998).

\bibitem{Durr+2:98b}
S. D\"{u}rr, T. Nonn,  and G. Rempe, 
\PRL\textbf{81}, 5705 (1998).

\bibitem{Mei+1:01}
M. Mei and M. Weitz, 
\PRL\textbf{86}, 559 (2001).

\bibitem{Durr:01}
S. D\"urr, 
\PRA\textbf{64}, 042113 (2001). 

\bibitem{Luis:01}
A. Luis,
\JPhysA\textbf{34}, 8597 (2001).

\bibitem{Zyczkowski+1:website}
K. \.Zyczkowski and W. Tadej,
http://chaos.if.uj.edu.pl/$\sim$karol/hadamard/

\bibitem{Tadej+1:06}
W. Tadej and K. \.Zyczkowski,  
\JournalTitle{Open Syst.\ Inf.\ Dyn.}\textbf{13}, 133 (2006).

\bibitem{Shannon:48}
C. E. Shannon,
\JournalTitle{Bell Sys.\ Tech.\ J.}\textbf{27}, 623 (1948).

\bibitem{Kulik+3:06}
E. V. Moreva, G. A. Maslennikov, S. S. Straupe, and S. P. Kulik,
\PRL\textbf{97}, 023602 (2006). 

\bibitem{HackDict}
http://www.worldwideschool.org/library/books/tech/computers/\linebreak[4]%
TheHackersDictionaryofComputerJargon/chap42.html

\bibitem{WeiHuiThesis}
W. H. Chee, \textit{Quantitative wave-particle duality in higher dimensions}
(unpublished thesis, Singapore, 2005).

\bibitem{Kurizki+4:06}
G. Kurizki, N. Bar-Gill, J. Clausen, M. Kol\'a\v{r}, and T. Opatrn\'y,
\JournalTitle{Quant.\ Inf.\ Processing} \textbf{5}, 463 (2006).

\bibitem{LQM-PP}
B.-G. Englert, 
\textit{Lectures on Quantum Mechanics --- Perturbed Evolution}
(World Scientific, Singapore, 2006).

\bibitem{Englert+2:01}
B.-G. Englert, C. Kurtsiefer, and H. Weinfurter,
\PRA\textbf{63}, 032303 (2001).

\bibitem{terHaar:67}
D. ter Haar, \textit{The Old Quantum Theory} 
(Pergamon Press, Oxford and New York, 1967).

\bibitem{Wheeler+1:83}
\textit{Quantum Theory and Measurement}, eds.\ J. A. Wheeler and
W. H. Zurek (Princeton University Press, Princeton, 1983).

\end{thebibliography}
\end{document}